\documentclass[a4paper]{article}

\usepackage[T1]{fontenc}
\usepackage{amssymb,amsmath,amsthm,graphicx}
\usepackage{dsfont}
\usepackage{algorithm}
\usepackage{algorithmic}
\usepackage{subcaption}
\usepackage{color}
\usepackage{multirow}

\newcommand{\blind}{0}

\newtheorem{rmk}{Remark}[section]

\begin{document}

\def\spacingset#1{\renewcommand{\baselinestretch}%
{#1}\small\normalsize} \spacingset{1}


\if0\blind
{
  \title{\bf Simultaneous Dimension Reduction and Clustering via the NMF-EM Algorithm}
  \author{L\'ena Carel\\
    Transdev and CREST, ENSAE, Universit\'e Paris Saclay\\
    and \\
    Pierre Alquier \\
    CREST, ENSAE, Universit\'e Paris Saclay}
  \maketitle
} \fi

\if1\blind
{
  \bigskip
  \bigskip
  \bigskip
  \begin{center}
    {\LARGE\bf Title}
\end{center}
  \medskip
} \fi

\bigskip
\begin{abstract}
Mixture models are among the most popular tools for clustering. However, when the dimension and the number of clusters is large, the estimation of the clusters become challenging, as well as their interpretation. Restriction on the parameters can be used to reduce the dimension. An example is given by mixture of factor analyzers for Gaussian mixtures. The extension of MFA to non-Gaussian mixtures is not straightforward. We propose a new constraint for parameters in non-Gaussian mixture model: the $K$ components parameters are combinations of elements from a small dictionary, say $H$ elements, with $H \ll K$. Including a nonnegative matrix factorization (NMF) in the EM algorithm allows us to simultaneously estimate the dictionary and the parameters of the mixture. We propose the acronym NMF-EM for this algorithm, implemented in the R package {\tt nmfem}. This original approach is motivated by passengers clustering from ticketing data: we apply NMF-EM to data from two Transdev public transport networks. In this case, the words are easily interpreted as typical slots in a timetable.
\end{abstract}
<
\noindent%
\vfill

\newpage
\spacingset{1.45} 

\section{Introduction}


With the growing ability to collect and store data in transports system, electricity consumption and more, urban computing is becoming a major tool in urban policy and planning~\cite{zheng2014urban}. For example, for transports system, there is a growing litterature on ticketing and smart-card data processing in trains and buses~\cite{morency2007measuring,pelletier2009smart,el2014understanding,poussevin2014mining,carel2017NMF,tonnelier2018}, bike-sharing systems~\cite{randriamanamihaga2013clustering,etienne2014model,bouveyron2015discriminative,hamon2015factorisation} or taxis~\cite{peng2012collective}. Our objective in this paper is to propose a clustering method for users, and for stations, that would be adapted to ticketing data collected by Transdev, a public network company. This method could be suitable for clustering structured high-dimensional data in other applications.

The range of machine learning and statistical tools used in urban computing is large. This goes from descriptive data-mining techniques as in~\cite{morency2007measuring} to statistical models as in~\cite{el2014understanding}. The model-based clustering approach in~\cite{el2014understanding} is actually close to our objective: journeys of a user are seen as realizations of multinomials random variables. The parameters of these distributions depends of the user only through the cluster the user belongs to. The complete model for journeys is thus a mixture of multinomials. The authors estimate the parameters and the clusters by the EM algorithm (see Chapter 9 in~\cite{bishop2007pattern} for an introduction; many R packages are available, {\tt mcclust}~\cite{scrucca2016mclust} is extremely complete for clustering with Gaussian mixtures, {\tt mixtools}~\cite{mixtools} is a more generalist package covering other distributions, including multinomials). Model-based clustering was also used for transport data in~\cite{etienne2014model,bouveyron2015discriminative} with nice results. However, there are some issues with this approach. When the dimension is large, the estimates are likely to have a large variance (curse of dimensionality). It might also be difficult to interpret clusters described by a huge number of parameters: it is indeed argued in~\cite{carel2017NMF} that some profiles in~\cite{el2014understanding} are not easily interpretable. It seems then necessary to reduce the dimension, that is, to impose some restrictions on the parameters that will reduce the variance and increase the interpretability.

Since the seminal work on model-based clustering~\cite{wolfe1963}, various examples of such restrictions have been proposed. We refer the reader to~\cite{fraley2002model,Bouveyron2014,mcnicholas2016,mcnicholas2016b,grun} for recent surveys on existing approaches (see also~\cite{mclachlan2004finite,handbook} for a more general overview on mixtures). A first approach is variable selection~\cite{raftery2006variable}. This method is now well understood from an empirical perspective~\cite{steinley2008selection} as well as from a theoretical point of view~\cite{maugis2009variable,maugis2009variableb}. See~\cite{celeux2017} for more recent advances and~\cite{fop2017variable} for a nice survey. The underlying assumption is that clusters differ only through a few variables. This assumption is satisfied in many examples presented in the aforementionned papers. However, it does not seem to be adapted to our case. The difference between two users, say a student and a retired person, is that the student has a regular travel schedule, while the retired person usually doesn't. This is a typical example of a strong structure that is not summarized by a small number of variables. Another approach for dimension reduction in mixtures is the mixture of factor analyzer (MFA) introduced in~\cite{ghahramani1996,mclachlan2003}, see~\cite{montanari2010,mcnicholas2008,murphy2017} for recent extensions. In MFA, the means and variances depends on the cluster, and the variance might be concentrated in some directions. This is more related to our objective, but this model was developped for mixture of Gaussians. Travels patterns are modeled by mixture of multinomials in~\cite{el2014understanding}.

In this paper, we propose a new model that can be seen as an adaptation of MFA to mixture of distributions with nonnegative parameters (including multinomial distributions). The decomposition in Gaussian factors in MFA is replaced by a nonnegative matrix factorization (NMF). Introduced by~\cite{lee1999learning}, NMF rewrites columns of a given matrix with nonnegative entries as combinations of elements in a small dictionary. These elements are often refered to as ``words''. These words play a somewhat similar role to factors in MFA, even though the formalism is different. For example these words are not modelled as random variables. We provide an adaptation of the celebrated EM algorithm to this setting. We refer to this algorithm as NMF-EM. It is available as an R package, {\tt nmfem}.

The paper is organized as follows. In Section~\ref{nmfem} we describe our model and the general form of NMF-EM. Motivated by the ticketing data, we provide the detailed form of the algorithm in the case of mixture of multinomials (Subsection~\ref{subsection-application-multinomial}). The clustering abilities of NMF-EM are compared to the ones of EM (without reduction of dimension) and of $k$-means in a short simulation study in Section~\ref{section-simulation-study}. We finally present results on ticketing data provided by the Transdev Group in Section~\ref{section-application-transdev} (more details on this real data study can be found in the supplementary material).

\section{Factorization of mixture parameters and the NMF-EM algorithm}
\label{nmfem}

\subsection{Factorization of mixture parameters}

Given a parametric family of distributions $(f_{\vartheta})_{\vartheta\in\mathbb{R}^M}$, assume the observations $Y_1,\dots,Y_n$ are i.i.d from
\begin{equation}
\label{equation:mixture:general}
\sum_{k=1}^K p_k f_{\theta_{\cdot,k}}(\cdot),
\end{equation}
where each $\theta_{\cdot,k} \in \mathbb{R}^M$ is a column of a $K\times M$ matrix $\theta$. For the sake of brevity, let $p=(p_1,\dots,p_K)$, which belongs to the simplex $\mathcal{S}_K=\{\rho\in\mathbb{R}_+^K: \rho_1+\dots+\rho_K = 1\}$. A way to rephrase this mixture model which is useful for clustering purposes is to introduce i.i.d hidden class variables: $Z_i=(Z_{i,1},\dots,Z_{i,K}) \sim \mathcal{M}ult(p,1)$. Here, $\mathcal{M}ult(p,1)$ denotes the multinomial distribution, that is, the probability that $Z_i$ is the $k$-th basis vector $(0,\dots,1,\dots,0)$ is given by $p_k$. Taking $Y_i \bigl| (Z_{i,k} = 1)\bigr. \sim f_{\theta_{\cdot,k}}(\cdot)$ implies that the $Y_i$'s are actually i.i.d from~\eqref{equation:mixture:general}.

In model-based clustering, estimation of the $Z_i$'s allow to assign each $Y_i$ to a cluster $k$ while the estimation of $\theta_{\cdot,k}$ provides a summary of the information on location, scale and shape of cluster $k$. Still, as argued in the introduction, when the dimension $M$ is too large, this information can be unreliable and uneasy to interpret. Many dimension reduction methods were proposed, among them MFA for mixtures of Gaussians. A standard mixture of Gaussian in $\mathbb{R}^d$ is $Y_i \bigl| (Z_{i,k} = 1)\bigr. \sim \mathcal{N}(\mu_k,\Sigma_k) $, the simplest form of MFA is given by $Y_i \bigl| (Z_{i,k} = 1)\bigr. \sim \mathcal{N}(\mu_k,\Lambda_k \Lambda_k^T + \sigma^2 I) $, where $\Lambda_k$ is a $d\times H$ matrix with $H\ll d$. Thus, the estimation of the $d\times d$ matrix $\Sigma_k$ is reduced to the estimation of the much smaller $H\times d$ matrix $\Lambda_k$. An interpretation of this model is that $Y_i$ depends not only on the hidden variable $Z_i$ but also on hidden factors $X_i\sim \mathcal{N}(0,I_H)$: $\mathbb{E}(Y_i|X_i=x,Z_{i,k}=1)=\Lambda_k x + \mu_k$. See the references given in the introduction, e.g Section 5 in~\cite{Bouveyron2014}. This model provides reduction of dimension and has a nice interpretation, but is is not direct to extend it beyond Gaussian variables.

In the case where of multinomial distributions, and more generally in the case where the parameters $\vartheta$ of $(f_\vartheta)_{\vartheta\in\mathbb{R}^M}$ are actually nonnegative, one could think of restrictions on the mixture parameters matrix $\theta$ that would similarly involve a small number $H$ of hidden factors. But one has to be careful: Gaussian hidden factors would in general not generate nonnegative parameters. In a celebrated paper~\cite{lee1999learning}, Lee and Seung proposed a dimension reduction tool for matrices with nonnegative entries: NMF (nonnegative matrix factorization). The idea is to factorize a $K\times M$ matrix $\theta$ as
\begin{equation}
\label{eqn-nmf}
\underbrace{\left(
       \begin{array}{c c c}
        \theta_{1,1} & \dots & \theta_{1,K} \\
        \vdots & \ddots & \vdots \\
        \theta_{M,1} & \dots & \theta_{M,K}
       \end{array}
       \right)}_{\theta}
       =
\underbrace{\left(
       \begin{array}{c c c}
        \Phi_{1,1} & \dots & \Phi_{1,H} \\
        \vdots & \ddots & \vdots \\
        \Phi_{M,1} & \dots & \Phi_{M,H}
       \end{array}
       \right)}_{\Phi}
\underbrace{\left(
       \begin{array}{c c c}
        \Lambda_{1,1} & \dots & \Lambda_{1,K} \\
        \vdots & \ddots & \vdots \\
        \Lambda_{H,1} & \dots & \Lambda_{H,K}
       \end{array}
       \right)}_{\Lambda}
\end{equation}
with $H\leq K,M$, under the assumption that all the entries in $\Phi$ and $\Lambda$ are nonnegative. When $H \ll K,M$, the dimension reduction is substantial. NMF rewrites columns of a given matrix as positive combinations of elements, or words, in a small dictionary $\Lambda$. It turns out that this dictionary is often easily interpretable. NMF was succesfully used as a data mining tool in document clustering~\cite{xu2003document,shahnaz2006document}, collaborative filtering and recommender systems on the Web~\cite{koren2009matrix,luo2014}, dictionary learning for images~\cite{lee1999learning},  topic extraction in texts~\cite{paisley2014bayesian} or time series recovering~\cite{mei2016recovering}, among others. It was also used as a data mining tool for transports data by~\cite{hamon2015factorisation,peng2012collective,poussevin2014mining,tonnelier2018} and our previous work~\cite{carel2017NMF}: we ``compressed'' the data $Y_1,\dots,Y_n$ using an NMF and then used a (model-free) clustering algorithm on the compressed observations. The improvement in terms of interpretability with respect to~\cite{el2014understanding} was substantial.
However, this approach was completely {\it ad hoc}: there are many possible criterion to approximate NMF: the Poisson-likekihood~\cite{lee1999learning,lee2001algorithms}, the quadratic criterion or Gaussian-likelihood~\cite{boyd2011distributed,lee2001algorithms}, the Ikuro-Saito divergence~\cite{fevotte2009nonnegative}... In a model-free approach, the choice of the criterion is difficult. The mixture model~\eqref{equation:mixture:general} leads to a natural criterion: the likelihood.

We are finally in position to define our model: we use NMF as a restriction on nonnegative parameters in mixture models. That is, $Y_1,\dots,Y_n$ are i.i.d from
\begin{equation}
\label{equation:mixture:positive}
g_{p,\Phi,\Lambda}(\cdot) = \sum_{k=1}^K p_k f_{(\Phi \Lambda)_{\cdot,k} }(\cdot)
\end{equation}
or equivalently, $Y_i|(Z_{i,k}=1)$ is drawn from $f_{(\Phi \Lambda)_{\cdot,k}}$ and $Z_i\sim\mathcal{M}ult(p,1)$. The model is parametrized by $p\in\mathcal{S}_K$, $\Lambda\in\mathbb{R}_+^{M\times H}$ and $\Phi\in\mathbb{R}_+^{H\times K}$. For short, put $Y=(Y_1,\dots,Y_n)$ and $Z=(Z_1,\dots,Z_n)$. The log-likelihood is given by
$$
\ell (\Phi,\Lambda,p|Y)  = \sum\limits_{i=1}^n \log \left ( \sum\limits_{k=1}^K  p_k f_{(\Phi \Lambda)_{\cdot,k} }(Y_i) \right ).
$$
This model can be seen as offering a connection between ``model-free clustering'' relying on NMF or spectral clustering as in~\cite{ding2005equivalence,yang2016low} and model-based clustering. Unrestricted mixture models can of course be seen as a special case by taking $H=K$ and $\Lambda = I_K$.

\begin{rmk}
The first example we have in mind is the mixture of multinomials that was used in~\cite{el2014understanding} to model travel patterns. As our main application, this example is detailed in Subsection~\ref{subsection-application-multinomial}. Note a similarity with the Latent Dirichlet Allocation (LDA) model in~\cite{blei2003latent}: LDA involves two layers of multinomials. First, a topic is a multinomial on words, then a text is described by a multinomial on topics. However, LDA does not involve clusters of similar texts. It was not designed as a clustering tool.

Beyond multinomials, any distribution with nonnegative parameters can be used. Consider sales analysis. Assume that the owner of a supermarket observes, for each good $m\in\{1,\dots,M\}$ and each customer $i\in\{1,\dots,n\}$, the number of items of $m$ bought by $i$ during one year: $Y_{i,m}$. Put $Y_i=(Y_{i,1},\dots,Y_{i,M})$. We propose the model $Y_{i,m}|(Z_{i,k}=1)\sim \mathcal{P}(\theta_{m,k})$, a Poisson distribution. The column $\theta_{\cdot,k}$ is the ``standard basket'' of any customer $i$ in cluster $k$. But the number of goods is so huge that the estimation of standard baskets is subject to a large variance, and prevents their interpretation. In~\eqref{eqn-nmf}, the columns of $\Phi$ are representations of columns of $\theta$ in a smaller subspace. It is likely that substituable goods are gathered. This example is simply a model-based version of the NMF analysis used in~\cite{koren2009matrix,luo2014}, see also~\cite{wu2007} for an early application on the Netflix prize data. Sales analysis, customer clustering and recommender systems are indeed applications of NMF that generated a huge number of publications. More examples could include exponential or gamma mixtures in survival analysis, or Pareto and Weibull mixtures in extreme analysis.
\end{rmk}

We now discuss the adaptation of the EM algorithm to this parameter restriction.

\subsection{The NMF-EM algorithm}

We remind the expression of the completed likelihood
$$
\ell (\Phi,\Lambda,p|Y,Z) = \sum\limits_{i=1}^n  \sum\limits_{k=1}^K Z_{i,k} \log \left (p_k f_{(\Phi \Lambda)_{\cdot,k} }(Y_i) \right ).
$$
A step of the EM algorithm, given current parameters $(\Phi^{(c)},\Lambda^{(c)},p^{(c)})$ is as follows:
\begin{eqnarray}
 \nonumber
 \text{{\bf E-step}: } Q^{(c)}(\Phi,\Lambda,p) &=& \mathbb{E}_{\Phi^{(c)},\Lambda^{(c)},p^{(c)}}[\ell(\Phi,\Lambda,p|Y,Z)\vert Y]\\
 \nonumber
&=& \sum\limits_{i=1}^n  \sum\limits_{k=1}^K \mathbb{E}_{\Phi^{(c)},\Lambda^{(c)},p^{(c)}}[Z_{i,k}|Y] \log \left (p_k f_{(\Phi \Lambda)_{\cdot,k} }(Y_i) \right ) \\
\text{ and }  t_{i,k}^{(c)} &:=& \mathbb{E}_{\Phi^{(c)},\Lambda^{(c)},p^{(c)}}[Z_{i,k}|Y]
    =\frac{p_k^{(c)} f_{(\Phi^{(c)}\Lambda^{(c)})_{\cdot,k}}(Y_i)}{\sum \limits_{k'=1}^K p_{k'}^{(c)}f_{(\Phi^{(c)}\Lambda^{(c)})_{\cdot,k'}}(Y_i)}.
   \label{tik}
\end{eqnarray}
\begin{equation}
 \text{{\bf M-step}: }
     \label{functionQ}
    (\Phi^{(c+1)},\Lambda^{(c+1)},p^{(c+1)}) := \underset{\Phi_{j,h},\Lambda_{h,k}\geq 0}{\arg\max} Q^{(c)}(\Phi,\Lambda,p).
\end{equation}
Obviously, the challenging step is the M-step. While we obviously have, for $k\in\{1,\dots,K\}$,
\begin{equation}
   \label{equa-p}
     p_k^{(c+1)} = \frac{\sum \limits_{i=1}^n t_{i,k}^{(c)}}{\sum \limits_{i=1}^n\sum \limits_{k'=1}^K t_{i,k'}^{(c)}},
\end{equation}
the nonnegativity constraint on $\Phi$ and $\Lambda$ makes the optimization with respect to these two matrices much harder. This is where one has to plug ideas from the NMF literature. Many options might be possible, depending on the form of $f_\vartheta(\cdot)$. The most commonly used algorithm is the so-called multiplicative update, an alternating optimization method with respect to $\Phi$ and $\Lambda$, that was proposed in the seminal papers~\cite{lee1999learning,lee2001algorithms}. Other algorithms include  ADMM~\cite{boyd2011distributed,sun2014alternating}, alternating projected gradient~\cite{Lin07}, and for Bayesian approaches, Monte-Carlo methods~\cite{paisley2014bayesian} and variational approximations~\cite{alquier2016oracle}. A numerical comparison of many algorithms can be found in~\cite{Lin07}. In practice, the multiplicative update is efficient in many settings and is very simple to use: it does not depend on any tuning parameter such as the step size in gradient based method. So this is the method we will use from now. This method iterates a step in $\Phi$, and a step in $\Lambda$. Each step is shown to improve the fit criterion in~\cite{lee2001algorithms}. Note that the author claims that it also leads to convergence, but as argued in~\cite{gonzalez2005accelerating} the proof of this fact is actually incomplete. We explicit the multiplicative update in the case of mixture of multinomials below.

\subsection{The NMF-EM algorithm for mixture of multinomials}
\label{subsection-application-multinomial}

\begin{figure}[!htp]
    \centering
    \includegraphics[scale=0.6]{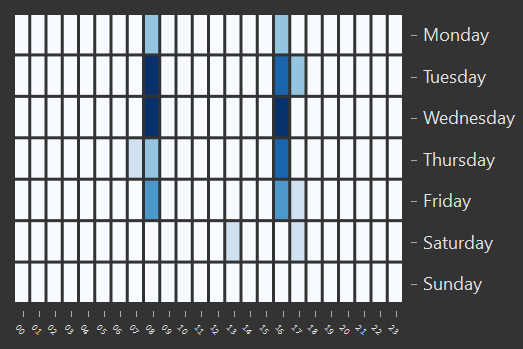}
    \caption{Temporal profile of a network user, taken from the data described in Section~\ref{section-application-transdev}. Opacity is proportional to the number of smart-card validations.}
    \label{fig:profil_user}
\end{figure}

In~\cite{el2014understanding} the authors modeled a passenger temporal profile by a mixture of multinomial distribution. The time and days of smart card validations of a passenger $i$ are recorded over a period of time (e.g. 1 month). The numbers of journeys, $N_i$, is not our variable of interest, and will be considered as deterministic. We obtain as a result a vector
$ Y_i = (Y_{i,1},\dots,Y_{i,M})^T \in \mathbb{R}^M $
where each coordinates represents the number of travels at a given pair time-day during the considered period. Note that of course $\sum_{k=1}^M Y_{i,k} = N_i$, let $N=\sum_{i=1}^n N_i$ be the total number of journeys. We consider a hourly grid, that is, Mon-12am, Mon-1am, etc... to Sun-11pm, with means that $M=7\times 24 = 168$. An example of a traveler profile is given in Figure~\ref{fig:profil_user}.

It is natural to assume that there are clusters of passengers with rather similar profiles: for examples, employees with similar work hours or students in the same University are likely to commute at similar times. We follow the previous construction: we define the hidden cluster variables, $Z_i \sim\mathcal{M}ult(p,1)$ for some $p\in\mathcal{S}_K$. We then set
$$ Y_i \bigl| (Z_{i,k}=1) \bigr. \sim \mathcal{M}ult(\theta_{\cdot,k},N_i) $$
where $\theta_{\cdot,k}\in\mathcal{S}_M$ is the $k$-th column of an $M\times K$ matrix $\theta$ that satisfies $\theta=\Phi\Lambda$ where $\Phi$ is $M\times H$ and $\Lambda$ is $H\times K$ for some $H\leq M,K$.
The log-likelihood is given by
$$
\ell(\Phi,\Lambda,p | Y)
    = \sum_{i=1}^n \log \left\{
    \sum_{k=1}^K p_k \left[
       N_i!
    \prod_{j=1}^M \frac{( \Phi \Lambda)_{j,k}^{Y_{i,j}}}{Y_{i,j}!}
     \right]
    \right\}.
$$
Note that a simple way to ensure $\theta_{\cdot,k}\in\mathcal{S}_M$ is to impose similar constraints on the columns of $\Phi$ and $\Lambda$. So we define $\mathcal{M}_{M,H,K}$ as the set of all pairs $(\Phi,\Lambda)$ of matrices $M\times H$ and $H\times K$ respectively, with $\Phi_{\cdot,k},\Lambda_{\cdot,j}\in \mathcal{S}_M $ for any $k$ and $j$. Note that we actually have $H(M-1)+K(H-1)+K-1$ degrees of freedom for the parameters of our model $(\Phi,\Lambda)\in\mathcal{M}_{M,H,K}$ and $p\in\mathcal{S}_K$. This knowledge is required for computing model selection criterion such as AIC (see the discussion on model selection in Subsection~\ref{subsectionHK} below).

Let $(\hat{\Phi},\hat{\Lambda},\hat{p})$ denote the MLE, that is, a maximizer of $\ell(\Phi,\Lambda,p | Y)$ with respect to $(\Phi,\Lambda)\in \mathcal{M}_{M,H,K}$ and $p\in\mathcal{S}_K$. We explicit the NMF-EM algorithm to approximate $(\hat{\Phi},\hat{\Lambda},\hat{p})$.

From \eqref{tik}, values $t_{i,k}^{(c)}$ are given by
$$
    t_{i,k}^{(c)} = \frac{p_{k}^{(c)} N_i!\prod \limits_{j=1}^{M}\frac{\left ( \sum \limits_{h=1}^{H}\Phi_{j,h}^{(c)}\Lambda_{h,k}^{(c)} \right )^{Y_{i,j}}}{Y_{i,j}!}}{\sum \limits_{k'=1}^{K} p_{k'}^{(c)}N_i!\prod \limits_{j=1}^{M}\frac{\left ( \sum \limits_{h=1}^{H}\Phi_{j,h}^{(c)}\Lambda_{h,k'}^{(c)} \right )^{Y_{i,j}}}{Y_{i,j}!}} 
    = \frac{p_{k}^{(c)} \prod \limits_{j=1}^{M}\left ( \sum \limits_{h=1}^{H}\Phi_{j,h}^{(c)}\Lambda_{h,k}^{(c)} \right )^{Y_{i,j}}}{\sum \limits_{k'=1}^{K} p_{k'}^{(c)}\prod \limits_{j=1}^{M}\left ( \sum \limits_{h=1}^{H}\Phi_{j,h}^{(c)}\Lambda_{h,k'}^{(c)} \right )^{Y_{i,j}}}.
$$
We have
\begin{align*}
    Q^{(c)}(\Phi,\Lambda,p) &= \sum\limits_{i=1}^n  \sum\limits_{k=1}^K t_{i,k}^{(c)} \log \left (p_k N_i!\prod \limits_{j=1}^{M}\frac{\left ( \sum \limits_{h=1}^{H}\Phi_{j,h}\Lambda_{h,k} \right  )^{Y_{i,j}}}{Y_{i,j}!} \right ) \\
    &= \sum\limits_{i=1}^n  \sum\limits_{k=1}^K t_{i,k}^{(c)} \Biggl[ \log(p_k) +\log(N_i!)
    \\
    & \qquad \qquad + \sum \limits_{j=1}^{M} \left ( Y_{i,j} \log \left ( \sum \limits_{h=1}^{H} \Phi_{j,h} \Lambda_{h,k} \right ) - \log(Y_{i,j}!)\right )  \Biggr].
\end{align*}
As stated in~\eqref{equa-p}, $
     p_k^{(c+1)} \propto \sum \limits_{i=1}^n t_{i,k}^{(c)} $,
and
$$
(\Phi^{(c+1)},\Lambda^{(c+1)}) = \underset{(\Phi,\lambda)\in\mathcal{M}_{M,H,K}}{\arg\max}
\sum_{k=1}^K 
\sum_{j=1}^M \left( \sum_{i=1}^n Y_{i,j} t_{i,k}^{(c)} \right) \log \left ( \sum \limits_{h=1}^{H} \Phi_{j,h} \Lambda_{h,k} \right ).
$$
Put $ M^{(c)}_{j,k} = \sum_{i=1}^n Y_{i,j} t_{i,k}^{(c)} $ for short. The previous equation becomes
\begin{equation}
\label{nmf-EM}
(\Phi^{(c+1)},\Lambda^{(c+1)}) = \underset{(\Phi,\Lambda)\in\mathcal{M}_{M,H,K}}{\arg\max}
\sum_{k=1}^K
\sum_{j=1}^M M^{(c)}_{j,k}  \log \left ( \sum \limits_{h=1}^{H} \Phi_{j,h} \Lambda_{h,k} \right ).
\end{equation}
The maximization in~\eqref{nmf-EM} is equivalent to the minimization of
\begin{equation}
D(M^{(c)}||\Phi \Lambda) := -
\sum_{k=1}^K 
\sum_{j=1}^M \left\{ M^{(c)}_{j,k} \log \left ( \sum \limits_{h=1}^{H} \Phi_{j,h} \Lambda_{h,k} \right ) - \sum \limits_{h=1}^{H} \Phi_{j,h} \Lambda_{h,k} \right\}.
\end{equation}
Indeed, for $(\Phi,\Lambda)\in\mathcal{M}_{M,H,K}$ we have
$
\sum_{k=1}^K \sum_{j=1}^M \sum_{h=1}^{H} \Phi_{j,h} \Lambda_{h,k} = \sum_{j=1}^M \sum_{k=1}^K (\Phi \Lambda)_{j,k}
 = \sum_{j=1}^M 1 = M
$
that does not depend on $(\Phi,\Lambda)$. The multiplicative algorithm in~\cite{lee2001algorithms} was actually introduced to minimize $D(M^{(c)}||\Phi \Lambda)$. So we just use the update steps of~\cite{lee2001algorithms} (steps 9 and 10 in Algorithm~\ref{algo:nmf:em} below) followed by a renormalization of the matrices $\Phi$ and $\Lambda$ in order to ensure that the columns remain in the parameter space (steps 10 and 12). This completes the derivation of the NMF-EM algorithm for mixture of multinomials: Algorithm~\ref{algo:nmf:em} page~\pageref{algo:nmf:em}. We implemented this algorithm for the R software~\cite{R1996}, the package {\tt nmfem} can be found on the CRAN repository.

\begin{algorithm}[h]
\caption{NMF-EM}
\label{algo:nmf:em}
\begin{algorithmic} [1]
\STATE Fix $\epsilon>0$. Choose arbitrary $\Phi^{(0)}$, $\Lambda^{(0)}$ and $p^{(0)}$; $c:=0$, $ {\rm CRIT} := \infty $.
\WHILE{$|\ell(\Phi^{(c)},\Lambda^{(c)},p^{(c)})-{\rm CRIT}| > \epsilon$}
\STATE $ {\rm CRIT} := \ell(\Phi^{(c)},\Lambda^{(c)},p^{(c)})$.
\STATE For all $i\in\{1,\dots,n\}$ and $k\in\{1,\dots,K\}$,
\begin{eqnarray*}
    t_{i,k}^{(c)} := \frac{p_{k}^{(c)} \prod \limits_{j=1}^{M}\left ( \sum \limits_{h=1}^{H}\Phi_{j,h}^{(c)}\Lambda_{h,k}^{(c)} \right )^{Y_{i,j}}}{\sum \limits_{k'=1}^{K} p_{k'}^{(c)}\prod \limits_{j=1}^{M}\left ( \sum \limits_{h=1}^{H}\Phi_{j,h}^{(c)}\Lambda_{h,k'}^{(c)} \right )^{Y_{i,j}}}
\text{ and }
   p_k^{(c+1)} =: \frac{\sum \limits_{i=1}^n t_{i,k}^{(c)}}{\sum \limits_{i=1}^n\sum \limits_{k'=1}^K t_{i,k'}^{(c)}}.
\end{eqnarray*}
\STATE $\forall j,k \quad M^{(c)}_{j,k} = \sum_{i=1}^n Y_{i,j} t_{i,k}^{(c)}$.
\STATE Initialization of $\Phi$ and $\Lambda$ (arbitrarily), $q:=\infty$.
\WHILE{$|Q^{(c)}(\Phi,\Lambda,p^{(c+1)})-q| > \epsilon$}
\STATE $q:=Q^{(c)}(\Phi,\Lambda,p^{(c+1)}) $.
\STATE $  \forall h,k \quad \Lambda_{h,k} \leftarrow  \Lambda_{h,k} \frac{\sum_{j} \Phi_{j,h} M^{(c)}_{j,k} / (\Phi \Lambda)_{j,k} }
 {\sum_{j} \Phi_{j,h}} $
\STATE $ \forall h,k \quad \Lambda_{h,k} \leftarrow \frac{\Lambda_{h,k}}{\sum_{k'} \Lambda_{h,k'}}$
\STATE $ \forall j,h \quad \Phi_{j,h}  \leftarrow \Phi_{j,h} \frac{ \sum_{k} \Lambda_{h,k} M^{(c)}_{j,k} / (\Phi \Lambda)_{j,k} }{\sum_{k} \Lambda_{h,k}} $
\STATE $ \forall j,h \quad \Phi_{j,h}  \leftarrow \frac{\Phi_{j,h}}{\sum_{h'} \Phi_{j,h'}}$
\ENDWHILE
\STATE $(\Phi^{(c+1)},\Lambda^{(c+1)}):= (\Phi,\Lambda)$.
\STATE $c:= c+1$.
\ENDWHILE
\end{algorithmic}
\end{algorithm}

\subsection{Discussion on the choice of $H$ and $K$}
\label{subsectionHK}

The choice of $K$ is not a straightforward issue in mixture models. {\it A fortiori} the choice of the pair $(H,K)$ is not easier.

From the likelihood and the degrees of freedom above we can derive the AIC and BIC criteron
\begin{align*}
{\rm AIC} & = \ell(\hat{\Phi},\hat{\Lambda},\hat{p}|Y) - \frac{H(M-1)+K(H-1)+K-1}{2}
\\
{\rm BIC} & = \ell(\hat{\Phi},\hat{\Lambda},\hat{p}|Y) - \frac{[H(M-1)+K(H-1)+K-1]\log(N)}{2}
\end{align*}
that are widely used in practice. Among the papers mentioned above, BIC is used for choosing the number of clusters of users in~\cite{bouveyron2015discriminative}. However, the consistency of AIC and BIC depend on conditions that might not be satisfied in mixture models. Other criteria more suitable for mixtures were investigated, like NEC and variants~\cite{biernacki1999improvement}. The slope heuristic~\cite{baudry2012} is known to give nice results in practice, and can also be show to be consistent in some settings~\cite{arlot2009data}. It is actually used in~\cite{el2014understanding} for mixtures of multinomials.

An important point is that our criterion should actually depend on the objective we have in mind. In regular models, AIC finds the optimal balance between bias and variance, while BIC identifies the true model, when there is one. These two objectives are usually not compatible~\cite{yang2005can}. In our collaboration with Transdev, interpretability of the results was actually one of the main objectives. We will use the slope heuristic in what follows.

\section{Simulation study}
\label{section-simulation-study}
In this section, we illustrate the dimension-reduction effect of NMF-EM on synthetic data. As our main interest is here clustering, we will compare the ``pairwise misclassification rate'' of NMF-EM with the one of EM and k-means algorithms -- that is, the proportion of pairs $(i,j)$ of individuals that are either assigned to the same component by the algorithm while they were actually generated from different components, or assigned to different components while they were simulated from the same.

The experimental setting is as follow: the dimension is $m=100$, for each experiment we generate $H_0$ words in $\mathbb{R}^m$ from a uniform distribution and then $K=10$ parameters $\theta_{\cdot,1},\dots,\theta_{\cdot,K}$ as linear combinations of these $H_0$ words - the coefficients of each parameters are independently drawn from a Dirichlet distribution $\mathcal{D}(\alpha,\dots,\alpha)$. We finally draw $n=1500$ individuals from the corresponding mixture of multinomials with uniform weights.

We compare NMF-EM with $H=4$, EM (without reduction of dimension) and k-means in various settings: in the case $H_0=4$, where the dimension reduction in NMF-EM is actually correct, and $H_0=8$ - this case is less favourable to NMF-EM with $H=4$ as it reduces too much the dimension... We also use different values for $\alpha$, leading to different shapes for the set of parameters $\{\theta_{\cdot,1},\dots,\theta_{\cdot,K}\}$. The results are in Tables \ref{tab:simulation_criteria_H4} and \ref{tab:simulation_criteria_H8}.

\begin{table}[!htbp]
\caption{Pairwise misclassification rate of the algorithms on simulated data when $H_0=4$ ($m = 100$, $n = 1500$, $N=150$, $K=10$).}
\begin{center}
\begin{tabular}{|r|c|c|c|c|c|c|c|}
\cline{2-8}
\multicolumn{1}{c|}{} & \texttt{$\alpha$ = .01} & \texttt{$\alpha$ = .1} & \texttt{$\alpha$ = .2} & \texttt{$\alpha$ = .3} & \texttt{$\alpha$ = .4} & \texttt{$\alpha$ = .5} & \texttt{$\alpha$ = .6}\\
 \hline
 \texttt{NMF-EM} & 9.5\% & 6.3\% & \textbf{4.7\%} & \textbf{5.0\%} & \textbf{4.9\%} & \textbf{5.3\%} & \textbf{5.9\%} \\
 \hline
 \texttt{EM} & 8.4\% & 6.8\% & 5.0\% & 5.7\% & 5.6\% & 5.9\% & 6.7\% \\
 \hline
 \texttt{k-means} & \textbf{6.4\%} & \textbf{5.9\%} & 5.3\% & 5.5\% & 5.6\% & 6.0\% & 6.2\% \\
 \hline
 \cline{2-8}
\multicolumn{1}{c|}{} & \texttt{$\alpha$ = .7} & \texttt{$\alpha$ = .8} & \texttt{$\alpha$ = .9} & \texttt{$\alpha$ = 1.0} & \texttt{$\alpha$ = 1.1} & \texttt{$\alpha$ = 1.2} & \texttt{$\alpha$ = 1.3} \\
 \hline
 \texttt{NMF-EM} & \textbf{6.5\%} & \textbf{6.7\%} & \textbf{6.7\%} & 7.6\% & 7.3\% & 7.5\% & 8.8\% \\
 \hline
 \texttt{EM} & 7.2\% & 7.3\% & 7.0\%  & 7.7\% & 8.1\% & 8.0\% & 8.9\% \\
 \hline
 \texttt{k-means} & 6.6\% & 6.7\% & 6.8\% & \textbf{7.0\%} & \textbf{7.2\%} & \textbf{7.1\%} & \textbf{7.5\%} \\
 \hline
\end{tabular}
\end{center}
\label{tab:simulation_criteria_H4}
\end{table}%

\begin{table}[!htbp]
\caption{Pairwise misclassification rate of the algorithms on simulated data when $H_0=8$ ($m = 100$, $n = 1500$, $N=150$, $K=12$).}
\begin{center}
\begin{tabular}{|r|c|c|c|c|c|c|c|}
\cline{2-8}
\multicolumn{1}{c|}{} & \texttt{$\alpha$ = .01} & \texttt{$\alpha$ = .1} & \texttt{$\alpha$ = .2} & \texttt{$\alpha$ = .3} & \texttt{$\alpha$ = .4} & \texttt{$\alpha$ = .5} & \texttt{$\alpha$ = .6}\\
 \hline
 \texttt{NMF-EM} & 5.2\% & 4.5\% & 5.8\% & 5.8\% & 6.5\% & 6.9\% & 8.1\% \\
 \hline
 \texttt{EM} & 4.3\% & 3.1\% & \textbf{3.1\%} & \textbf{3.8\%} & 5.0\% & 6.1\% & 6.1\% \\
 \hline
 \texttt{k-means} & \textbf{3.8\%} & \textbf{3.1\%} & 3.4\% & 4.0\% & \textbf{4.8\%} & \textbf{5.5\%} & \textbf{5.6\%} \\
 \hline
 \cline{2-8}
\multicolumn{1}{c|}{} & \texttt{$\alpha$ = .7} & \texttt{$\alpha$ = .8} & \texttt{$\alpha$ = .9} & \texttt{$\alpha$ = 1.0} & \texttt{$\alpha$ = 1.1} & \texttt{$\alpha$ = 1.2} & \texttt{$\alpha$ = 1.3} \\
 \hline
 \texttt{NMF-EM} & 8.2\% & 9.1\% & 10.0\% & 10.5\% & 10.3\% & 11.3\% & 11.5\% \\
 \hline
 \texttt{EM} & 6.4\% & 7.2\% & 7.5\% & 7.5\% & 8.3\% & 8.6\% & 8.5\% \\
 \hline
 \texttt{k-means} & \textbf{5.8\%} & \textbf{6.3\%} & \textbf{6.3\%} & \textbf{6.5\%} & \textbf{6.8\%} & \textbf{7.0\%} & \textbf{6.9\%} \\
 \hline
\end{tabular}
\end{center}
\label{tab:simulation_criteria_H8}
\end{table}


So, when the intrinsic dimension is small enough, NMF-EM really improves the clustering ability of EM. In any case, our main claim is that it leads to easily interpretable clusters, a fact that will be illustrated in the next section.

\section{Application to ticketing data}
\label{section-application-transdev}

\subsection{Description of the data}
\label{subsection-description-data}
The data used in our study are the validations made during the month of September 2015 on one Transdev network in a medium size city. Ticketing data are the information obtained at each transaction made by a smart card on a validator system. For privacy reasons it is not possible to connect each validation to the user who made it. The feature that allows us to realize our study and create temporal profiles is a card number which is encrypted, and re-initialized every three months. It is thus impossible to follow the long-term behaviour of a user. This is the reason why we focus on a one month period. This period (September) have been chosen because it has no vacation nor bank holiday. During September 2015, more than $4,000,000$ check-ins have been made on the network by $232,430$ passengers. 

The data are agregated so that for each traveler, for each day of the week (Monday to Sunday) and each hour ($00$ to $23$), we have the number of validation during the studied period. A passenger profile is thus defined by $24*7=168$ features. Figure~\ref{fig:profil_user} page~\pageref{fig:profil_user} already provided an example of a temporal profile of one of the users. This traveler uses mainly the network at $8$ a.m and $4$ p.m.


\begin{rmk}
We used the same strategy to create stations profiles: for each station, for each day of the week and each hour of the day, we know the number of validations that occured at this station during the study period. In Figure~\ref{fig:profil_station}, we show the temporal profile of the station ``Palais de Justice'' (courthouse), a tramway station in the city center. This station has travelers all day long, but knows an attendance peak every day from $4$ to $6$ p.m.
\begin{figure}[!htp]
    \centering
    \includegraphics[scale=0.6]{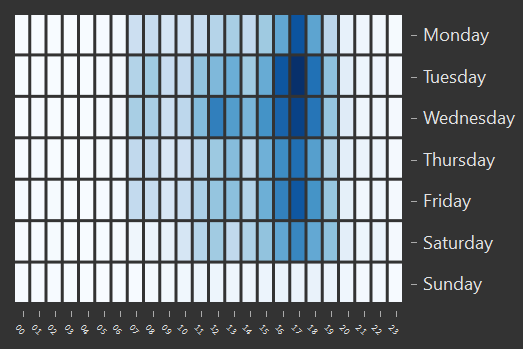}
    \caption{Temporal profile of station ``Palais de Justice''.}
    \label{fig:profil_station}
\end{figure}
The results of this analysis are provided in the supplementary material.
\end{rmk}

In order to avoid users who would not use their smart card enough to exhibit a clear pattern, data have been cleaned. We define a ``regular card holder'' as a card holder who
\begin{itemize}
 \item travelled on at least four days during September 2015 (so in particular we have $N_i\geq 4$);
 \item made their first boarding after 4 a.m each day at the same station $50\%$ of the time.
\end{itemize}
We only kept regular card holders for our analysis. After this cleaning step, we end up with $72,359$ profiles of passengers, which represent a bit more than $3,000,000$ check-ins -- that means $31\%$ of passengers represent $75\%$ of check-ins. We also have $475$ stations profiles. These data are provided in the {\tt nmfem} package.

\subsection{Passenger profile clustering}
\label{ssec:passengersclustering}
We first focus on passengers profiles clustering. This allows us to create groups of people that have similar temporal habits. The method used to create these clusters is the NMF-EM algorithm from Subection~\ref{subsection-application-multinomial}.

\begin{figure}[!htp]
    \centering
    \includegraphics[scale=0.4]{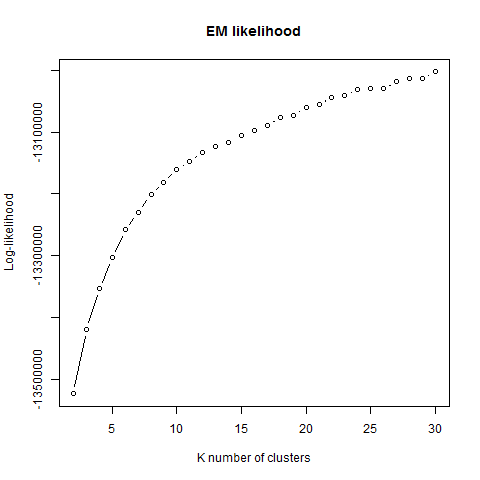}
    \caption{The log-likelihood as a function of $K$ under $H=K$.}
    \label{fig:llh_EM}
\end{figure}

To choose the parameters $H$ and $K$, we begin with the analysis of the log-likelihood of our model when $H=K$ for $K=2\dots30$. Note that the estimation of the model in this case can be made by the usual EM algorithm for multinomial mixture model. Figure \ref{fig:llh_EM} shows the evolution of the log-likelihood as a function of $K$.
This function clearly exhibits a linear behavior when $K\geq 10$. Thus, the slope heuristic suggests considering $K=10$.

\begin{figure}[!htp]
    \centering
    \includegraphics[scale=0.4]{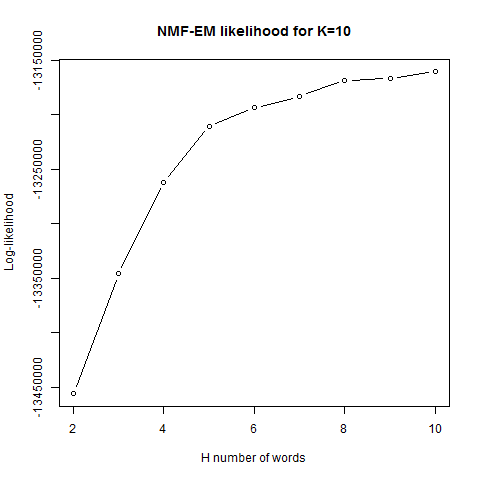}
    \caption{The log-likelihood as a function of $H\in\{2,\dots,K\} $ under $K=10$.}
    \label{fig:llh_K10}
\end{figure}

Now keeping $K=10$ fixed, we chose the value of $H$ in the same way. First, we plot the log-likelihood as a function $H$ in Figure \ref{fig:llh_K10}.
By using again the slope heuristic method, we choose $H=5$.

\begin{figure}[!htp]
    \centering
    \includegraphics[scale=0.4]{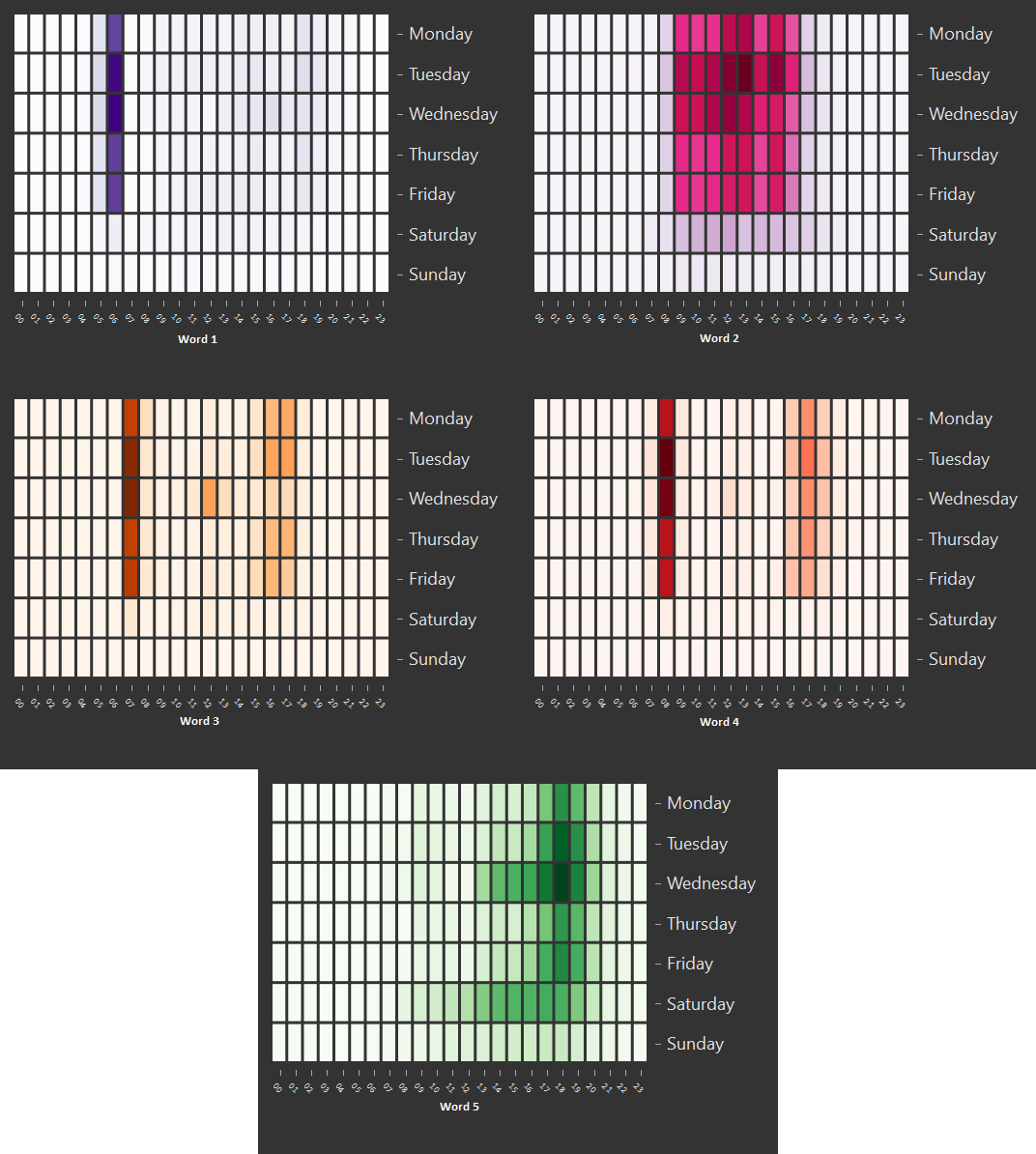}
    \caption{Words obtained by NMF-EM on users data with $K=10$ and $H=5$.}
    \label{fig:words}
\end{figure}

\begin{figure}[!htp]
    \centering
    \includegraphics[scale=0.6]{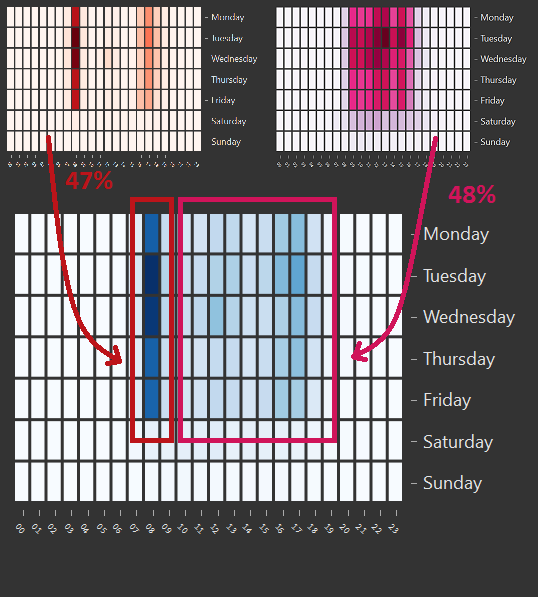}
    \caption{Decomposition of cluster $5$ from words $4$ and $2$.}
    \label{fig:decomposition}
\end{figure}

\begin{figure}[!htp]
    \centering
    \includegraphics[scale=0.4]{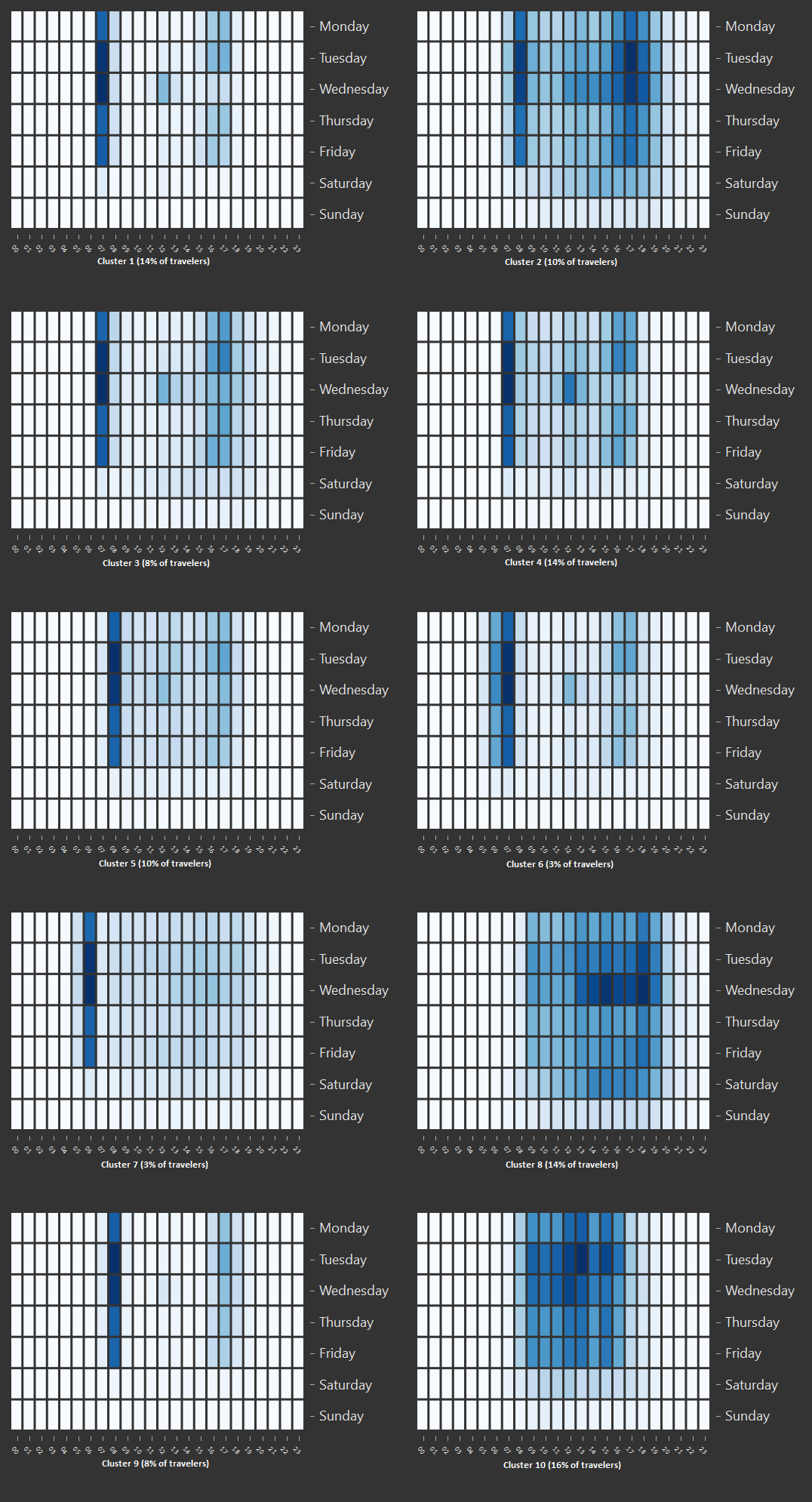}
    \caption{Clusters obtained by NMF-EM on users data with $K=10$ and $H=5$.}
    \label{fig:clusters}
\end{figure}

The $H=5$ words and the $K=10$ clusters are represented in Figure~\ref{fig:words} and in Figure \ref{fig:clusters} respectively. Remember that each cluster can be decomposed as a convex combination of words, some of them might have a null weight. For example, Figure~\ref{fig:decomposition} shows how the parameter of Cluster 5 can be written as a convex combination of words 4 and 2.

The interpretation of the words is direct:
\begin{enumerate}
 \item Word 1: travels between 6 a.m and 7 a.m.
 \item Word 2: diffuse component during off-peak periods (i.e. from 9 a.m to 4 p.m).
 \item Word 3: travels at school hours. Indeed it is composed of travel between 7 and 8 a.m and between 4 and 5 p.m, except on Wednesdays, when the afternoon travel is replaced by one at noon.
 \item Word 4: travels between 8 and 9 a.m.
 \item Word 5: late afternoon peak, from 5 to 7 p.m, and Wednesdays and Saturdays afternoon.
\end{enumerate}

We now attempt an interpretation of the clusters:
\begin{enumerate}
    \item Clusters $1$, $3$, $4$ and $6$ present high travel probabilities in the morning and in the afternoon except Wednesdays where the afternoon travel is replaced by a higher probability of travel around noon. These four clusters are typical of French schools and high-schools hours. The main differences are:
    \begin{enumerate}
        \item Cluster $1$: travels at 7 a.m and around 4 or 5 p.m.
        \item Cluster $3$: travel a bit more at 8 a.m.
        \item Cluster $4$: travelers are less susceptible to travel after 5 p.m.
        \item Cluster $6$: travels at 6 and 7 a.m.
    \end{enumerate}
    \item Cluster $5$: travels at 8 a.m and at 4 or 5 p.m.
    \item Cluster $7$: travels mainly at 6 a.m.
    \item Cluster $9$: travels at 8 a.m and at 5 p.m.
    \item Clusters $2$, $8$ and $10$: diffuse travel habits.
    \begin{enumerate}
        \item Cluster $2$: travels Mondays to Saturdays from 7 a.m to 7 p.m with highest probabilities at 8 a.m and 5 p.m Mondays to Fridays.
        \item Cluster $8$: diffuse travels Mondays to Saturdays from 9 a.m to 7 p.m.
        \item Cluster $10$: travels Mondays to Fridays from 9 a.m to 4 p.m.
    \end{enumerate}
\end{enumerate}


As a conclusion, NMF-EM provides clusters of users that are easily interpretable. In the supplementary material, we show how users profiles are related to demographic and economic variables.

\section{Conclusion}

We provided a new approach for dimension reduction that can be compared to MFA in non-Gaussian mixture models. This approach is based on NMF, an extremely popular data mining algorithm. We adapted the EM algorithm to this setting. This new algorithm, NMF-EM, is implemented in a package for R in the case of mixtures of multinomials. Results on simulated and real data are promising. In addition to a theoretical study of algorithm, future work should include an application to mixture of other distribution with nonnegative parameters like the Poisson distribution.

\noindent {\bf Acknowledgements}. We would like to thank the anonymous Referees and the Associate Editor for their constructive comments and suggestions.  We also thank Denis COUTROT and Nadir MEZIANI from Transdev for their support and comments on previous versions of this work.

\bigskip
\begin{center}
{\large\bf SUPPLEMENTARY MATERIAL}
\end{center}

\begin{description}

\item[Appendix] Contains the analysis of the clusters of users in Section~\ref{section-application-transdev}. We also apply NMF-EM on stations profile, and fully analyze another network (in the Netherlands).

\end{description}

\bibliographystyle{abbrv}

\newpage

\section*{Appendix}

\subsection*{Analysis of the clusters of users}

As written above, we have no personal information in our data. Therefore, we are not able to describe individually the users in each cluster. However, for each transaction made, we have the encrypted card number and the transport ticket used. So we can recover for each card the most used transport ticket during the period. This provides interesting information as some schemes are associated to age ranges (Young, Senior...) and to time periods (Unit, Annual or Monthly Subscription). Let us now provide the description of each cluster in terms of age ranges (Figures~\ref{fig:analyseAge_AD} to \ref{fig:analyseAge_GS} in Table \ref{tab:analyseAge}).

\begin{table}[!htb]
\begin{tabular}{cc}
\begin{subfigure}{0.45\textwidth}\centering\includegraphics[width=\columnwidth]{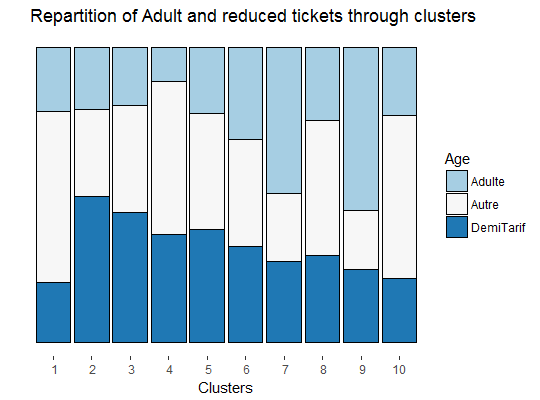}\caption{Adults (``Adult'') and Reduced tickets (``DemiTarif'')}\label{fig:analyseAge_AD}\end{subfigure}&
\begin{subfigure}{0.45\textwidth}\centering\includegraphics[width=\columnwidth]{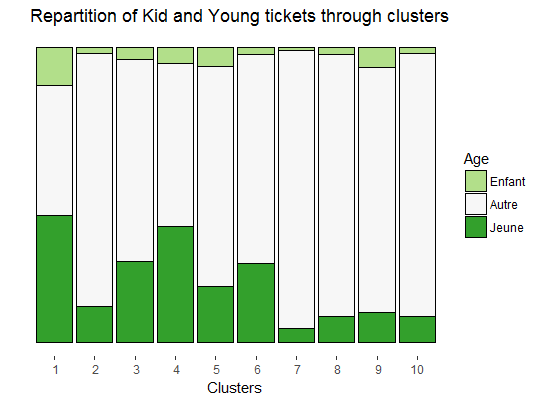}\caption{Children (``Enfant'') and Youngs (``Jeune'')}\label{fig:analyseAge_JE}\end{subfigure}\\
\newline
\begin{subfigure}{0.45\textwidth}\centering\includegraphics[width=\columnwidth]{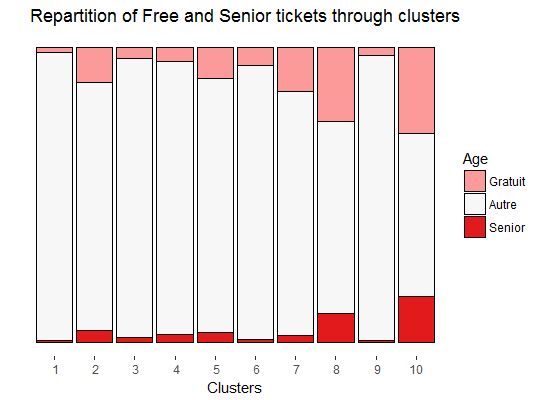}\caption{Free travelers (``Gratuit'') and Elderly (``Senior'')}\label{fig:analyseAge_GS}\end{subfigure}&
\\
\end{tabular}
\caption{Age range analysis of the clusters}
\label{tab:analyseAge}
\end{table}

Adults are more present in clusters $7$ and $9$, that are clusters with check-ins mostly in the morning. People benefiting from half-price are present in every cluster but with highest rates in clusters $2$, $3$, $4$ and $5$. Children (4 to 6) are not very present on the network, but they are more represented in clusters $1$, $5$ and $9$. Young travelers (6 to 25) are more present in clusters $1$ and $4$. These clusters correspond to scholar time slot. In clusters $8$ and $10$ there are large rate of seniors and free travelers. As these clusters have profiles of diffuse travels during the week and as free travelers are unemployed or low salaries people, these regroupments make sense. 

\begin{figure}[!htp]
    \centering
    \includegraphics[width=0.5\textwidth]{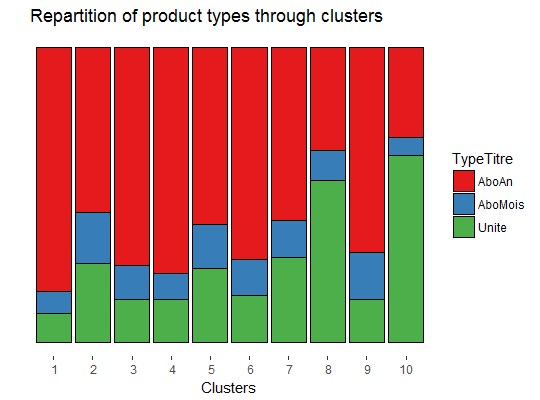}
    \caption{Transportation ticket type analysis of the clusters.}
    \label{fig:analyseTitre}
\end{figure}

Figure \ref{fig:analyseTitre} shows the repartition of transport ticket type through clusters.
Unit products are more used in clusters $8$ and $10$ that are clusters with a lots of seniors and free travelers. As they don't have obligations, they likely use unit products for occasional trips. Clusters $1$, $3$, $4$ and $9$, that have mostly scholar profiles althought have a large majority of annual subscripters. A possible interpretation is that schoolchildren and students are public transportation captives, and have to use the network in order to go to class every day. Thus, buying an annual pass is more advantageous than buying any other product type.

As described in Subsection~\ref{subsection-description-data}, we kept only users whose first trip of the day is made at the same station at least $50\%$ of the study time. That main ``morning station'' is thus called the ``home station'' as it gives us an estimation of the residence place of users. In Tables~\ref{tab:users_geolocation1} and~\ref{tab:users_geolocation2}, we can observe the shares of clusters by home stations. It shows the share of travelers identified as belonging to every cluster leaving near each station.
\begin{table}[!htb]
\begin{center}
\begin{tabular}{cc}
\begin{subfigure}{0.30\textwidth}\centering\includegraphics[width=\columnwidth]{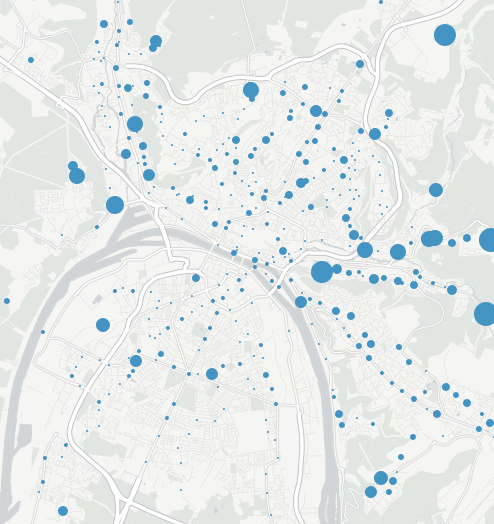}\caption{Cluster $1$}\label{fig:geo_cluster1}\end{subfigure}&
\begin{subfigure}{0.30\textwidth}\centering\includegraphics[width=\columnwidth]{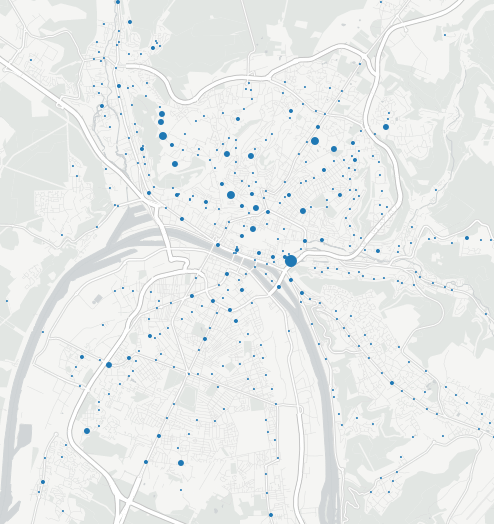}\caption{Cluster $2$}\label{fig:geo_cluster2}\end{subfigure}\\
\newline
\begin{subfigure}{0.30\textwidth}\centering\includegraphics[width=\columnwidth]{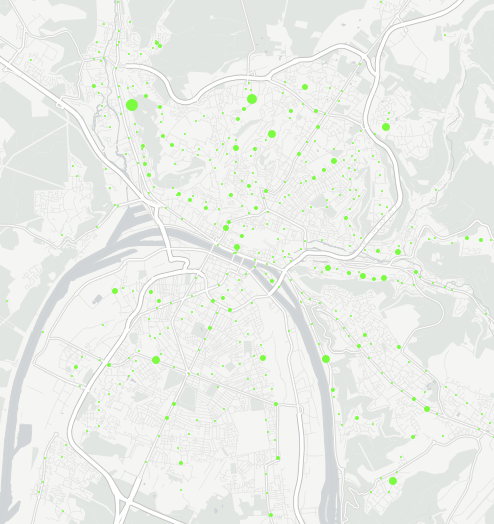}\caption{Cluster $3$}\label{fig:geo_cluster3}\end{subfigure}&
\begin{subfigure}{0.30\textwidth}\centering\includegraphics[width=\columnwidth]{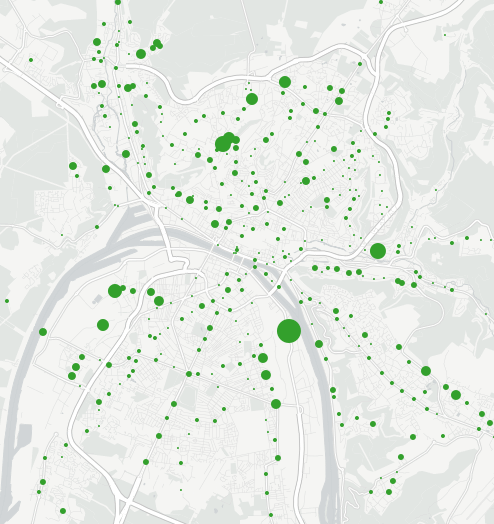}\caption{Cluster $4$}\label{fig:geo_cluster4}\end{subfigure}\\
\newline
\begin{subfigure}{0.30\textwidth}\centering\includegraphics[width=\columnwidth]{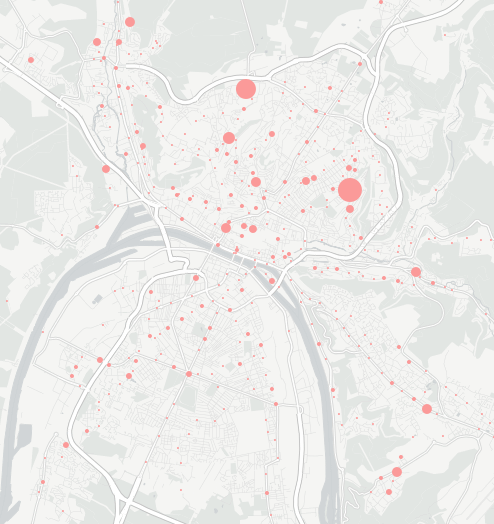}\caption{Cluster $5$}\label{fig:geo_cluster5}\end{subfigure}&
\begin{subfigure}{0.30\textwidth}\centering\includegraphics[width=\columnwidth]{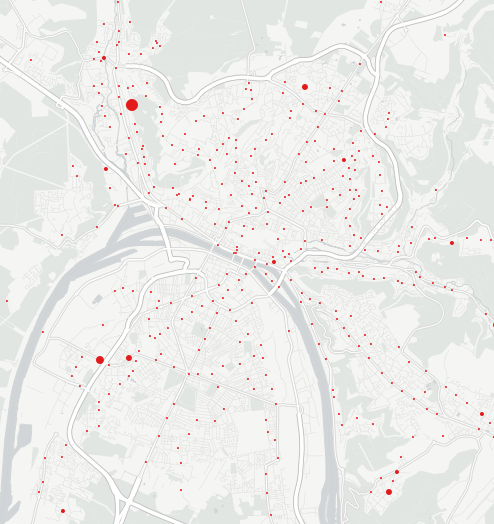}\caption{Cluster $6$}\label{fig:geo_cluster6}\end{subfigure}\\
\end{tabular}
\caption{Share of clusters per home station --- Clusters $1$ to $6$}
\label{tab:users_geolocation1}
\end{center}
\end{table}
\begin{table}[!htb]
\begin{center}
\begin{tabular}{cc}
\begin{subfigure}{0.30\textwidth}\centering\includegraphics[width=\columnwidth]{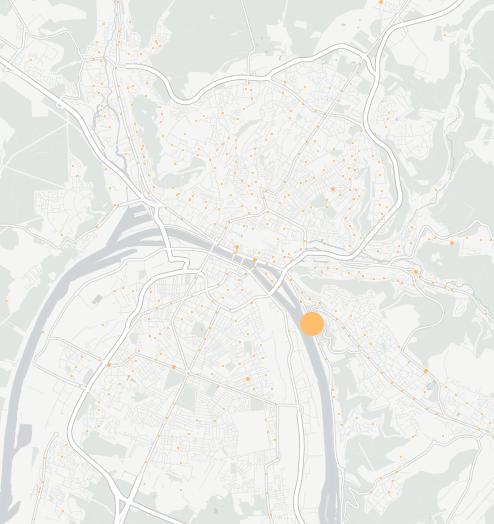}\caption{Cluster $7$}\label{fig:geo_cluster7}\end{subfigure}&
\begin{subfigure}{0.30\textwidth}\centering\includegraphics[width=\columnwidth]{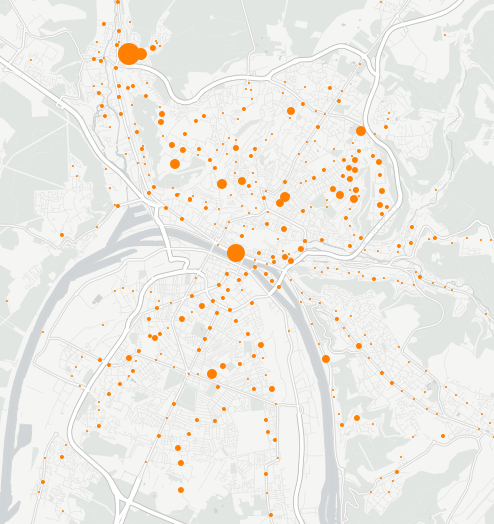}\caption{Cluster $8$}\label{fig:geo_cluster8}\end{subfigure}\\
\newline
\begin{subfigure}{0.30\textwidth}\centering\includegraphics[width=\columnwidth]{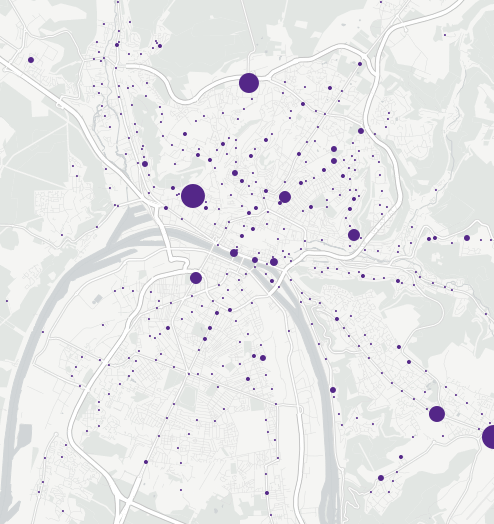}\caption{Cluster $9$}\label{fig:geo_cluster9}\end{subfigure}&
\begin{subfigure}{0.30\textwidth}\centering\includegraphics[width=\columnwidth]{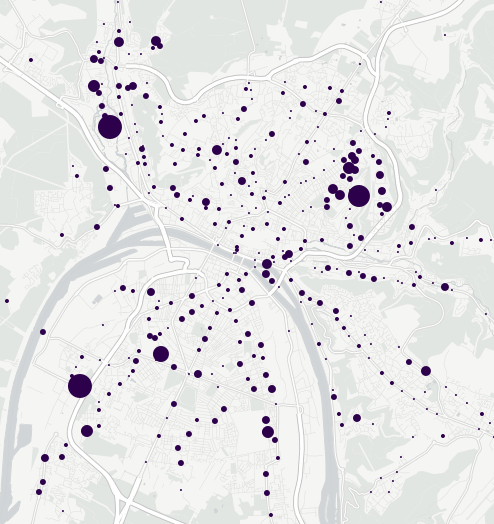}\caption{Cluster $10$}\label{fig:geo_cluster10}\end{subfigure}\\
\end{tabular}
\caption{Share of clusters per home station --- Clusters $7$ to $10$}
\label{tab:users_geolocation2}
\end{center}
\end{table}

We note that:
\begin{enumerate}
\item Cluster $1$: travelers are over represented at peripheral stations.
\item Cluster $2$: no particular pattern observed.
\item Cluster $3$: no particular pattern observed.
\item Cluster $4$: few stations show over representation of cluster $4$.
\item Cluster $5$: over representation of the cluster at two stations in the north.
\item Cluster $6$: no particular pattern observed.
\item Cluster $7$: One station is $100\%$ represented by cluster $7$. As only one user is assigned to this station, no particular pattern is observed.
\item Cluster $8$: the cluster is over represented at one station in the city center and at another further.
\item Cluster $9$: cluster $9$ is over represented in few stations in the center.
\item Cluster $10$: cluster is over represented in poorest neighborhoods of the city.
\end{enumerate}

\subsection*{Stations profile clustering}

Clustering the different stations of the network would allow us to better know the different type of stations, and to group them by temporal similarity. As we have very few number of stations ($475$), it is not safe to process as described above for the users clustering. Indeed, a $K$ larger than $6$ or $7$ leads to very small clusters. In place we fixed $H$ and $K$ {\it a priori} to $3$ and $5$ respectively.

\begin{figure}[!htp]
    \centering
    \includegraphics[scale=0.4]{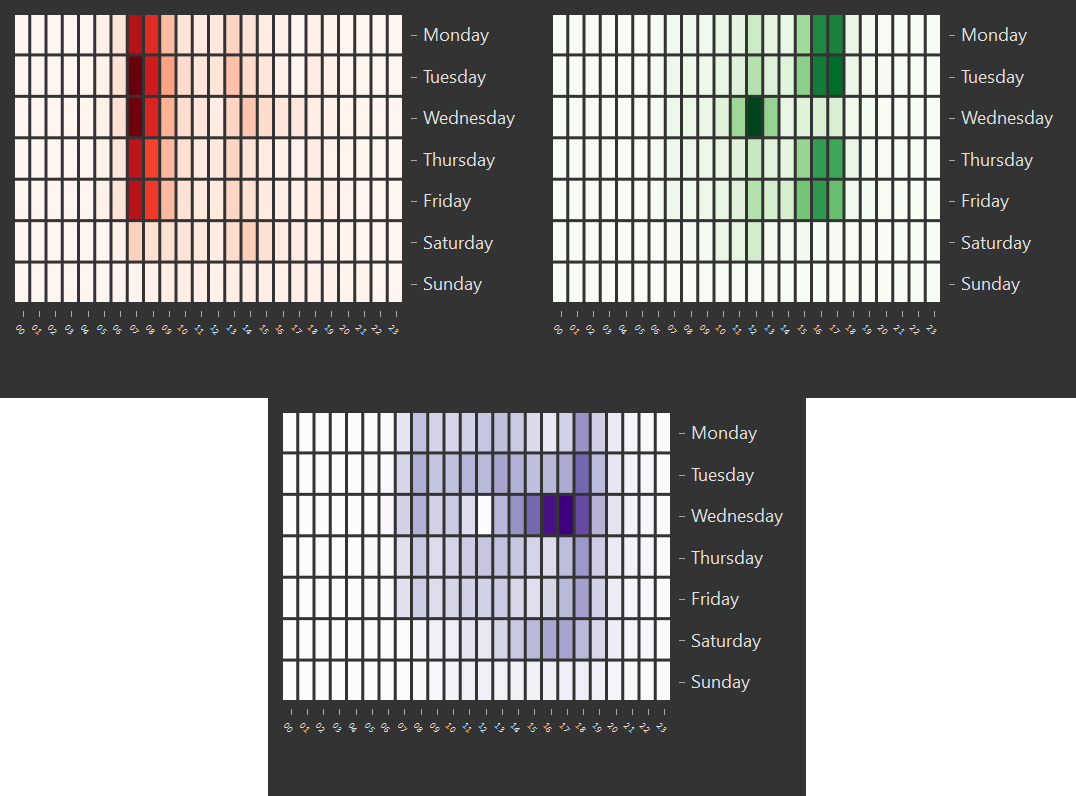}
    \caption{Words obtained by NMF-EM on stations data with $K=5$ and $H=3$.}
    \label{fig:words_station}
\end{figure}

The $3$ words obtained are the ones in Figure~\ref{fig:words_station}. 
The first time component is described by check-ins at 7 and 8 a.m. We will call it the ``morning component''. The second time component shows check-ins at 4 and 5 p.m on Mondays, Tuesdays, Thursdays and Fridays and check-ins at 12 p.m on Wednesdays. We will name it the ``end of school component''. The third component shows check-ins at 6 p.m, during Wednesdays afternoons, during Saturdays and off-peaks periods. This component will be called the ``off-peak component''.

\begin{figure}[!htp]
    \centering
    \includegraphics[scale=0.4]{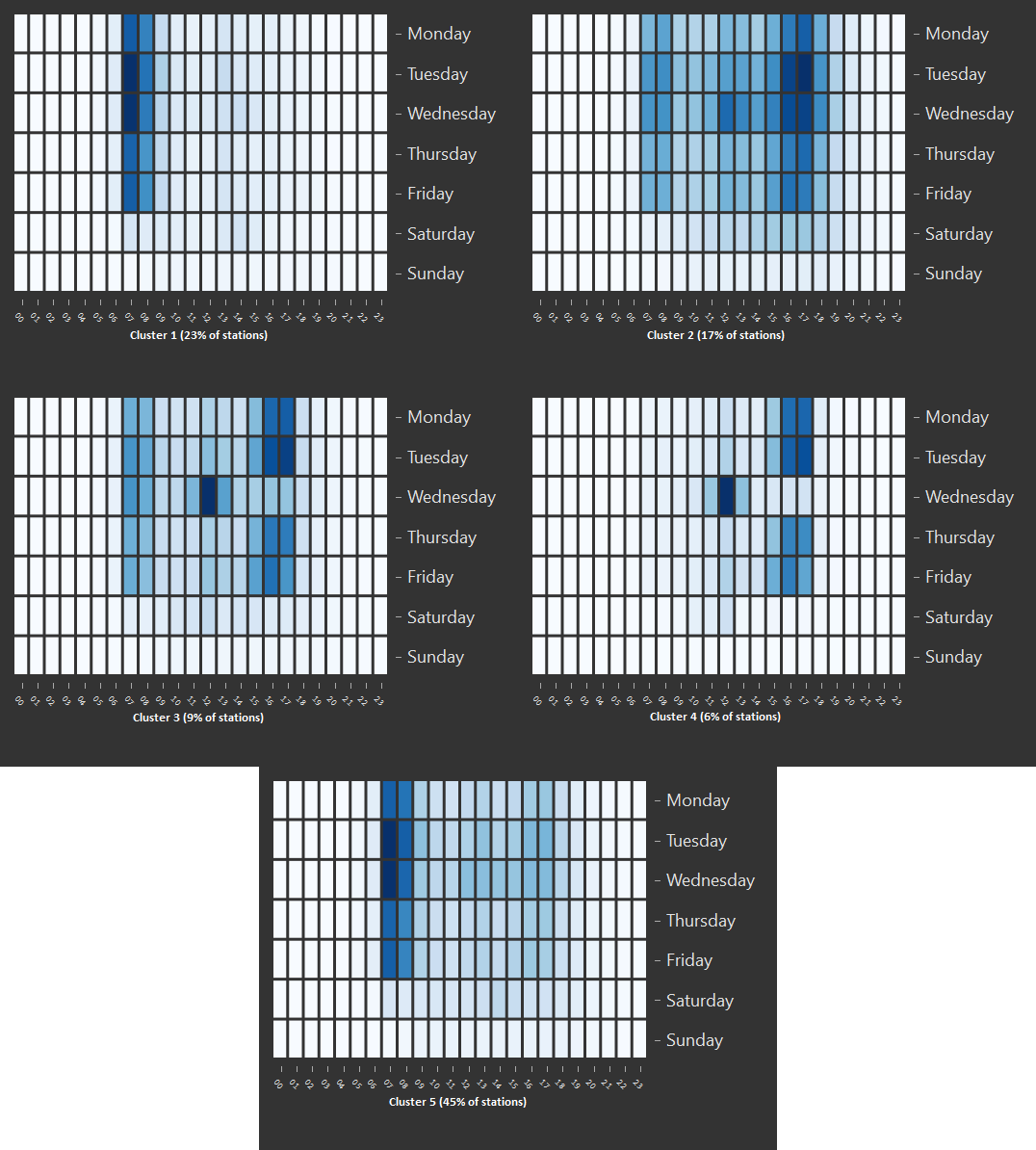}
    \caption{Clusters obtained by NMF-EM on stations data with $K=5$ and $H=3$.}
    \label{fig:clusters_station}
\end{figure}

Figure \ref{fig:clusters_station} shows the $5$ clusters.
Stations in cluster $1$ are stations where there are check-ins only in the morning at 7 or 8 a.m. These stations are likely in residential areas. In cluster $2$, the stations have check-ins all day long, but with highest probabilities during peaks. Stations in cluster $3$ have check-ins in the morning and at the end of school. They are likely to be near schools in residentials area. Stations in cluster $4$ have check-ins only at end of school times. Thus, these stations are probably near schools. Finally, stations in cluster $5$ are pretty similar than the ones in cluster $1$: a large majority of check-ins are made in the morning (7 or 8 p.m). The only difference is that it is more likely to have check-ins during the rest of the day in cluster $5$ than in cluster $1$.


Thanks to the French National Institute of Statistics and Economis Studies (INSEE), there are open data permitting us to introduce contextual information. Firstly, a database containing socioeconomic data on a grid of $200 m \times 200 m$ is available. We used two indicator of it: the number of inhabitants and the percentage of households living in collective housing per tiles. Secondly, we used a database referencing and geolocating every french company or administration. In this way, we were able to know the number of employees per tile. By clustering the tiles in the study area, we obtained different group of areas that will allow us to lead the study on stations more finely. Table~\ref{tab:tilesclusters} contains the description of the mean tile by cluster.

\begin{table}[!htb]
\centering
\caption{Description of tiles clusters}
\label{tab:tilesclusters}
\begin{tabular}{cccc}
  \hline
 \multirow{2}{*}{Tiles cluster} & \multirow{2}{*}{Inhabitants} & Percentage of & \multirow{2}{*}{Employees} \\ 
  & & collective housing & \\
  \hline
   1 & 223.75 & 57.73 & 824.43 \\ 
     2 & 162.03 & 30.08 & 40.32 \\ 
     3 & 268.74 & 58.66 & 2758.97 \\ 
     4 & 114.50 & 98.98 & 11576.50 \\ 
   \hline
\end{tabular}
\end{table}

As tiles contained in cluster $1$ and $2$, are those with the least number of employees, they can be described as residential areas.  Moreover, the percentage of collective housing allows to distinguish them. Indeed, cluster $1$ have more households living in collective housing than cluster $3$. That is why we will refer as tiles from cluster $1$ as residential areas in collective housing and as residential areas in individual housing for tiles from cluster $2$. Since the number of inhabitants and of employees are high, tiles from cluster $3$ will be refered as mixed areas. Finally, as the number of employees in cluster $4$ is very large, we will refer these tiles as business areas.

The figures in Table~\ref{tab:geo_clusters} show the geographical repartition of the five clusters. In Figure~\ref{fig:map_cluster1}, we oserve the stations contained in cluster $1$. This cluster groups stations that have check-ins only in the morning. On the figure, we observe that these stations are distant from the city center and are mainly located in residential areas. Figure~\ref{fig:map_cluster2} shows stations of cluster $2$, that have check-ins all day long with stronger attendance during peak-periods. These stations are mainly located in the city center. Figures~\ref{fig:map_cluster3} and~\ref{fig:map_cluster4} look alike. Indeed, clusters $3$ and $4$ have the ``end of school'' component and the points on the map are close to educational establishment. Figure~\ref{fig:map_cluster5} shows stations from cluster $5$. These stations have check-ins all day long but most are made in the morning. By looking at the map, we cannot notice any significant pattern.
\begin{center}
\begin{table}[!htb]
\begin{tabular}{cc}
\begin{subfigure}{0.45\textwidth}\centering\includegraphics[width=\columnwidth]{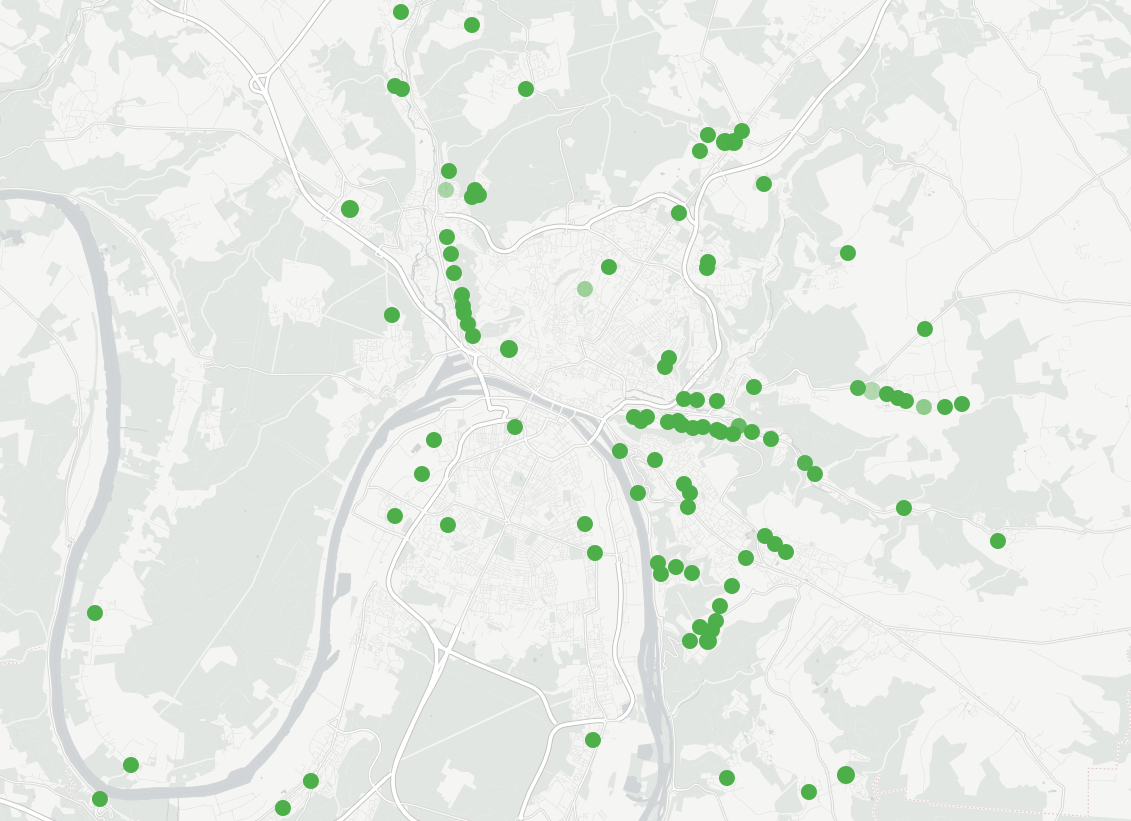}\caption{Cluster $1$}\label{fig:map_cluster1}\end{subfigure}&
\begin{subfigure}{0.45\textwidth}\centering\includegraphics[width=\columnwidth]{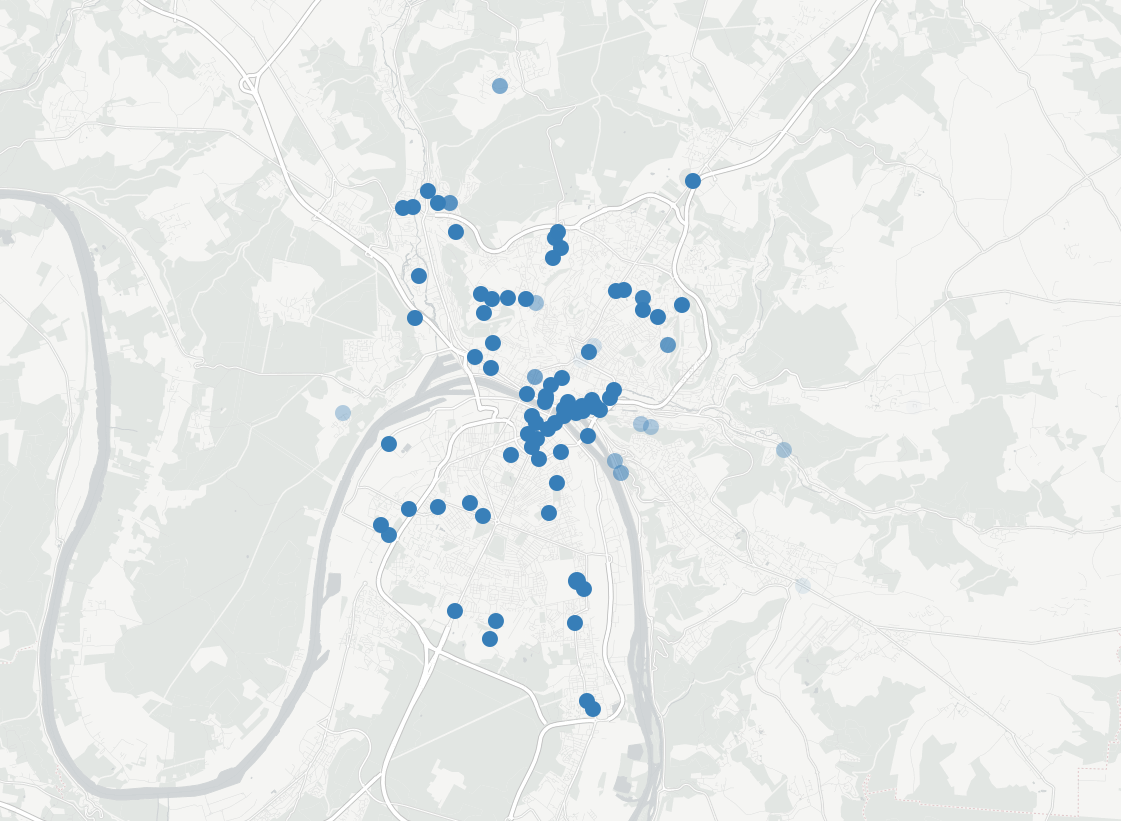}\caption{Cluster $2$}\label{fig:map_cluster2}\end{subfigure}\\
\newline
\begin{subfigure}{0.45\textwidth}\centering\includegraphics[width=\columnwidth]{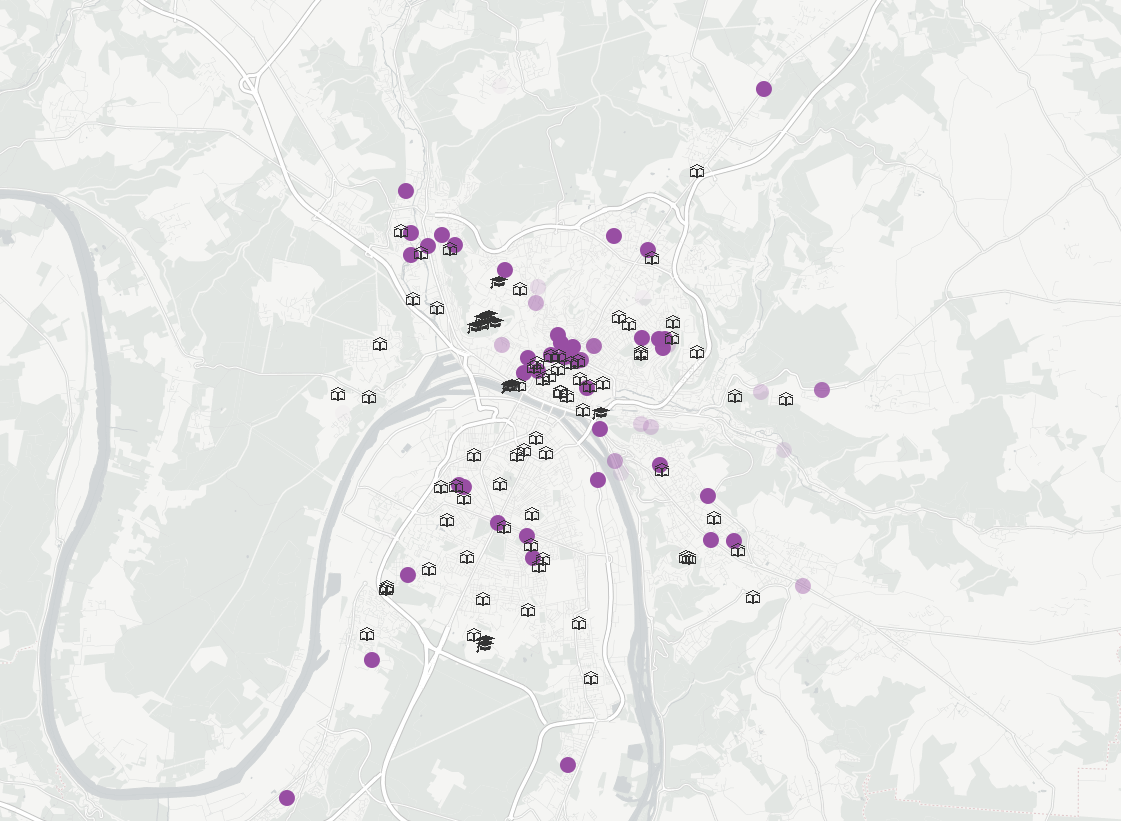}\caption{Cluster $3$}\label{fig:map_cluster3}\end{subfigure}&
\begin{subfigure}{0.45\textwidth}\centering\includegraphics[width=\columnwidth]{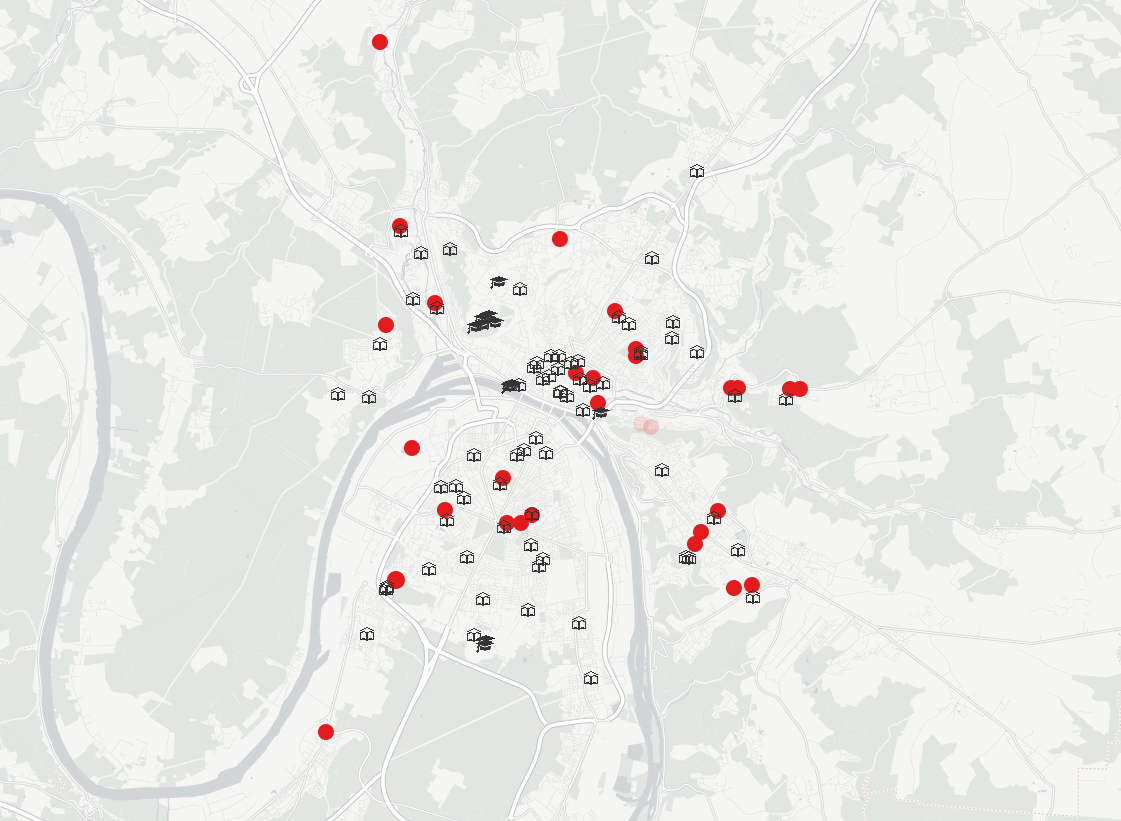}\caption{Cluster $4$}\label{fig:map_cluster4}\end{subfigure}\\
\newline
\begin{subfigure}{0.45\textwidth}\centering\includegraphics[width=\columnwidth]{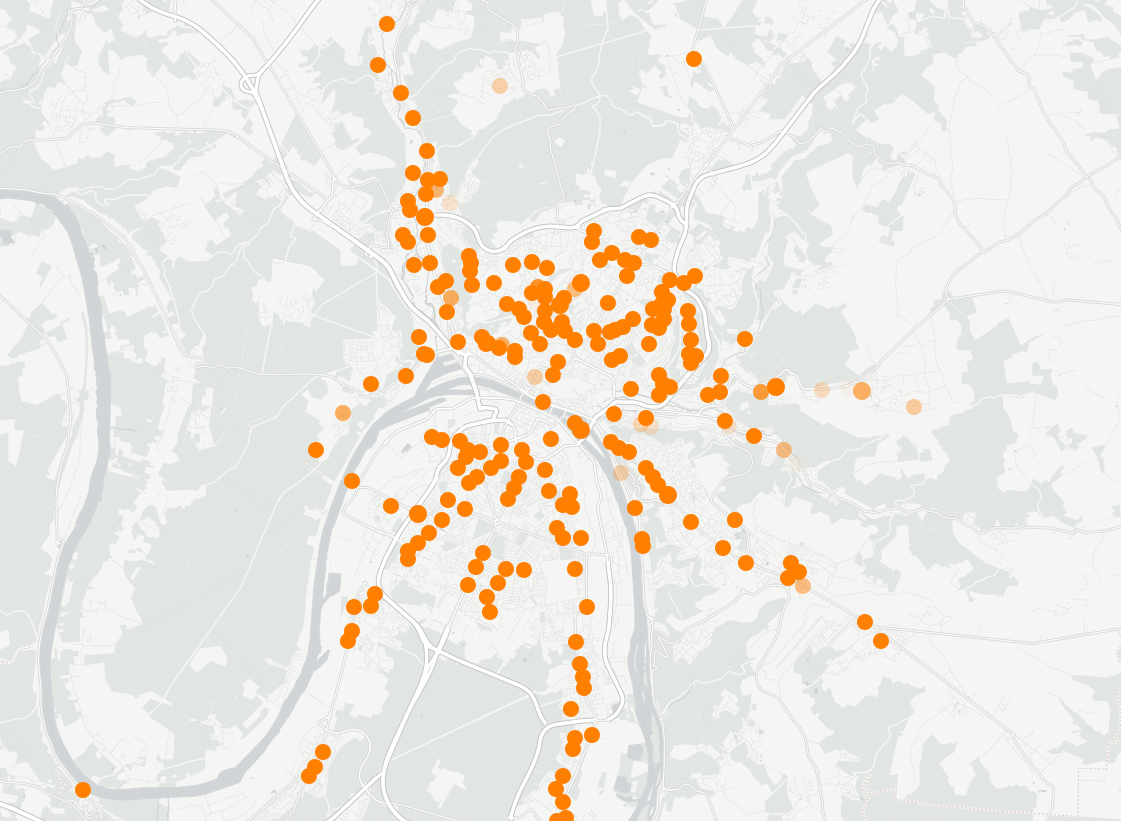}\caption{Cluster $5$}\label{fig:map_cluster5}\end{subfigure}&
\\
\end{tabular}
\caption{Map of the stations --- opacity of the points are proportional to the adequacy between the stations and the clusters.}
\label{tab:geo_clusters}
\end{table}
\end{center}

\subsection*{Passengers profile clustering on another network}
\begin{figure}[!htp]
    \centering
    \includegraphics[scale=0.4]{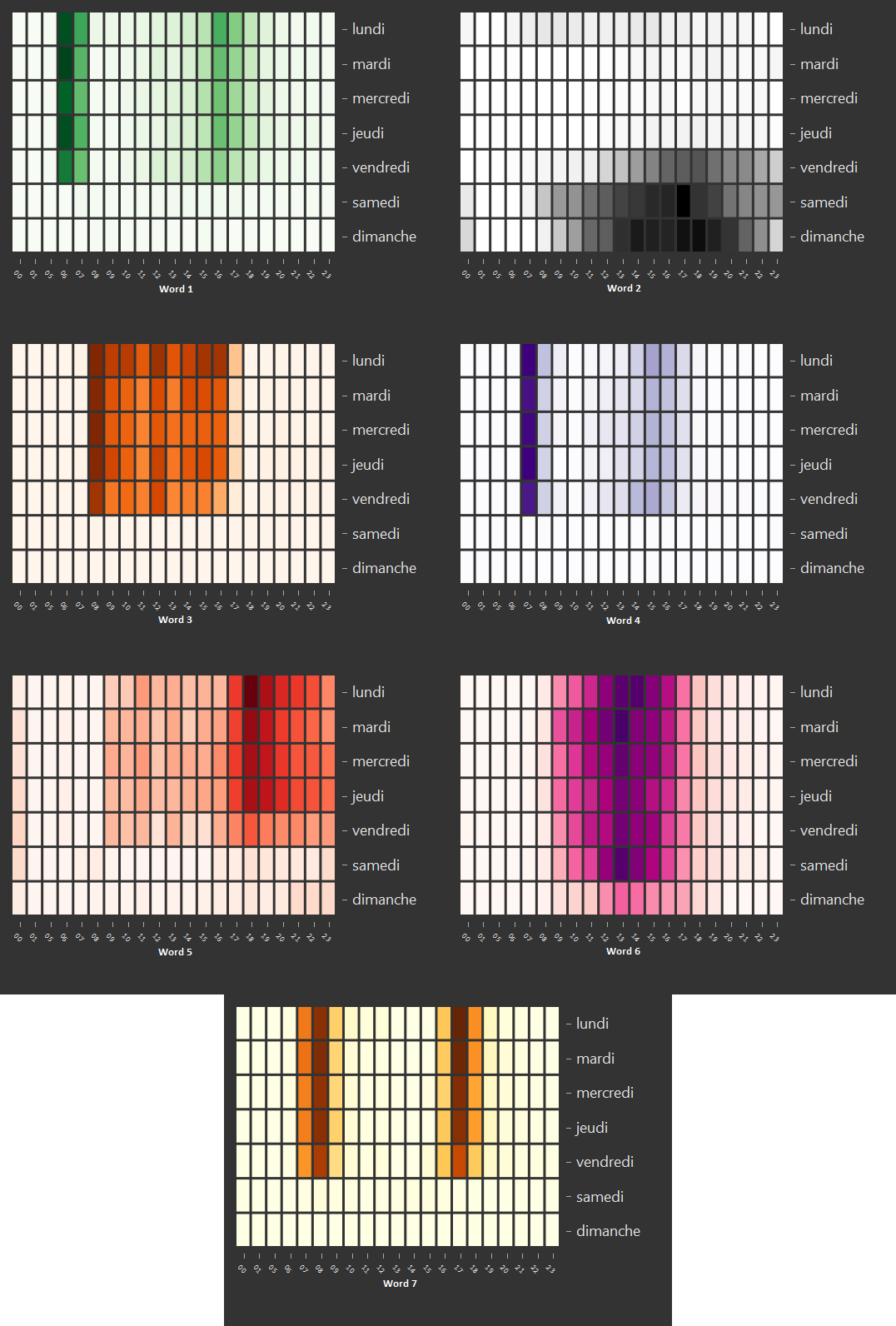}
    \caption{Words obtained by NMF-EM on users data with $K=10$ and $H=7$.}
    \label{fig:NL_words}
\end{figure}

\begin{figure}[!htp]
    \centering
    \includegraphics[scale=0.4]{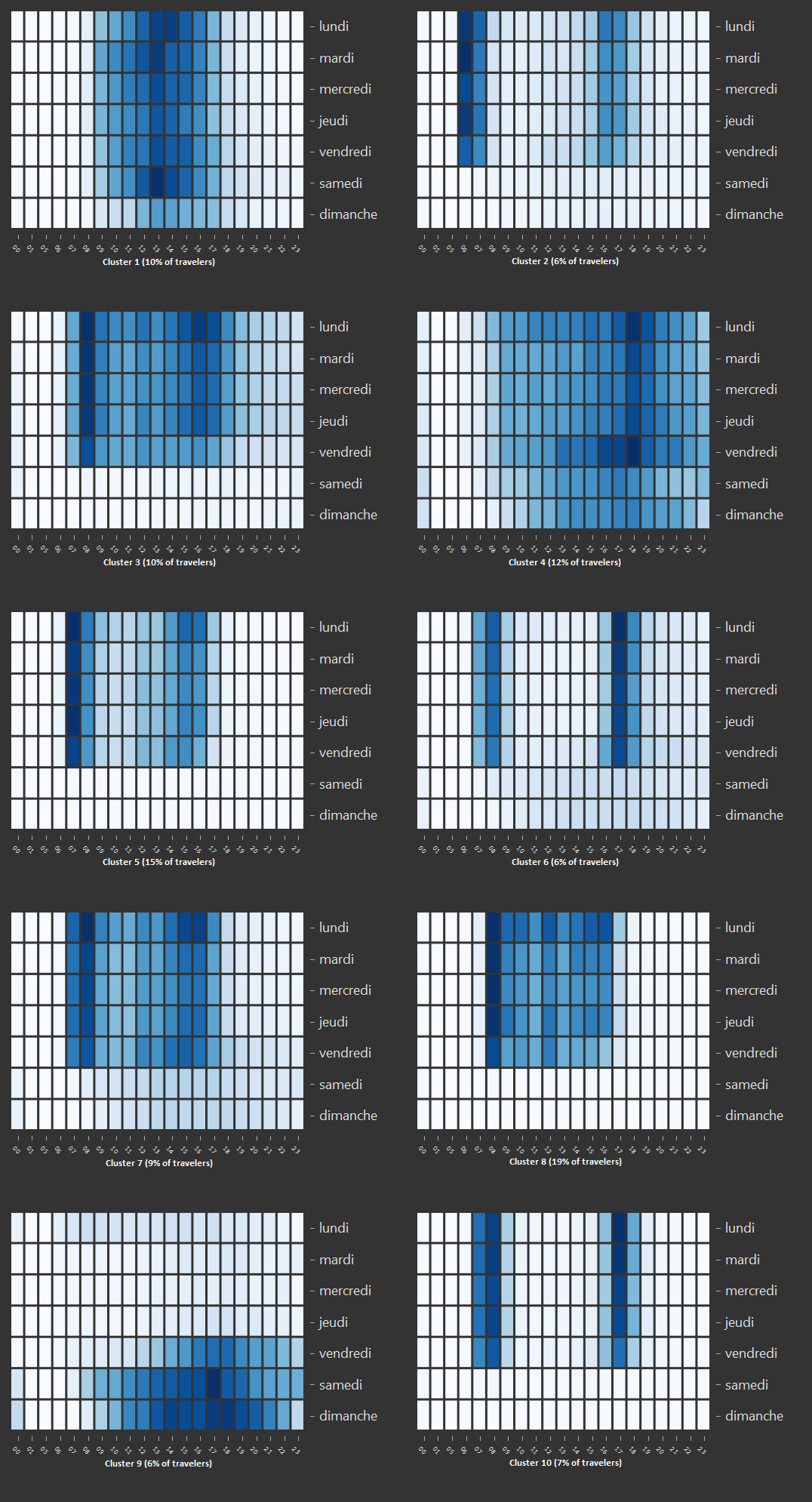}
    \caption{Clusters obtained by NMF-EM on users data with $K=10$ and $H=7$.}
    \label{fig:NL_clusters}
\end{figure}

To ensure efficiency of the algorithm, we applied it on another network located in the Netherlands. By applying the same model selection method as in Section~\ref{ssec:passengersclustering}, we obtained the optimal values of $K=10$ and $H=7$. Figures~\ref{fig:NL_words} and~\ref{fig:NL_clusters} contain respectively the profiles of the words and clusters obtained.

The interpretation of the words is:
\begin{enumerate}
    \item Word 1: travels at 6 or 7 a.m and slightly around 4 p.m during the week.
    \item Word 2: travels during the week-end.
    \item Word 3: diffuse travel habits from 8 a.m to 4 p.m Mondays to Fridays.
    \item Word 4: travels at 7a.m on weekdays.
    \item Word 5: diffuse habits with highest probabilities from 5 p.m to 12 a.m during the week.
    \item Word 6: diffuse habits from 9 a.m to 5 p.m with highest probability at 1 p.m Mondays to Saturdays.
    \item Word 7: travels at 8 a.m and 5 p.m.
\end{enumerate}

We can interpret the cluster as follows:
\begin{enumerate}
    \item Cluster 1: diffuse habits from 9 a.m to 5 p.m with highest probability at 1 p.m Mondays to Saturdays.
    \item Cluster 2: travels at 6 or 7 a.m and at 4 or 5 p.m during the week.
    \item Cluster 3: diffuse habits from 7 a.m to 6 p.m on weekdays.
    \item Cluster 4: diffuse travel habits from 9 a.m to 11 p.m.
    \item Cluster 5: travels at 7 or 8 a.m diffuse habits during the afternoon.
    \item Cluster 6: travels at 8 a.m and 5 p.m.
    \item Cluster 7: diffuse travel habits from 7 a.m to 5 p.m Mondays to Fridays.
    \item Cluster 8: diffuse habits from 8 a.m to 4 p.m during the week.
    \item Cluster 9: travels during the week-end.
    \item Cluster 10: travels at 7 or 8 a.m and around 4 p.m.
\end{enumerate}

\end{document}